\documentclass[epj]{svjour}
\usepackage{graphicx}
\usepackage{comment}
\usepackage{dcolumn}
\usepackage{bm}
\usepackage{todonotes}
\usepackage{amsmath}
\usepackage{amsfonts}
\usepackage{amssymb}

\usepackage{xcolor}

\usepackage[style=nature, citestyle=numeric-comp, backend=biber,sorting=none]{biblatex}
\title{Analytical solutions to intensity calculation in scintillation detectors and their application to the scintillation flash coordinates reconstruction}
\author{Khamitov Timur\inst{1,2} \and Vovchenko Ivan\inst{1,3,*}
}                     

\institute{Moscow Institute of Physics and Technology (National Research University), Dolgoprudnyi, Moscow region, 141700 Russia\and Institute for Nuclear Research (INR) of the Russian Academy of Sciences,   Moscow, 117312, Russia \and
Kotelnikov Institute of Radioengineering and Electronics of the Russian Academy of Sciences , 11-7 Mokhovaya, Moscow, 125009, Russia\\
\\
$^*$vovchenko@phystech.edu}

\date{Received: date / Revised version: date}
\abstract{
Detectors with continuous scintillator and matrix of photomultipliers attached are widely used in modern particle physics,  and various medical tomographs. 
There are two common ways to calculate the coordinates of the scintillation flash via photomultipliers energy outcome: Anger’s method and Monte Carlo simulations. 
In this article, we derive an analytical solution for the computation of the energy outcome from several photomultipliers built into the bottom face of the scintillation camera with a continuous scintillator shaped as a rectangular parallelepiped. 
The lateral faces of the scintillation camera are considered specular reflectors with reflective indexes $\sim 1$ or $0$. 
The analytical solution is used as a reference point for the scintillation flash coordinates reconstruction. 
The achieved results are similar to those the Monte Carlo simulation provides but require less computational time.
\PACS{
      {29.40.Mc}{Scintillation detectors}   \and
      {29.40.Gx}{Position-sensitive detectors}\and
      {07.85.Fv, 29.40.-n}{Gamma-ray detectors} \and
      {42.15.-i}{Geometrical optics}
     } 
} 

\begin{document}

\maketitle
%

\maketitle


\section{Introduction}
There are two basic types of position-sensitive scintillation detectors in experimental particle physics (including commercial medical SPECT and PET devices \cite{SPECT_AND_PET,moskal2016time,moskal2021simulating,beltrame2011ax,vandenberghe2020state,zhang2017quantitative,cherry2018total,madsen2007recent,khalil2011molecular,bailey2013evidence}): continuous scintillator with photomultipliers attached and pixel arrays. 
Both of them are meant to solve two tasks: determine the energy of incoming particle (which is proportional to the induced flash energy) and the coordinates of the scintillation flash. 
Pixel arrays can achieve better resolution and do not require any coordinate reconstruction methods, while detector with continuous scintillator does. Continuous detectors are cheaper for mass production. These two types of detectors can be combined \cite{two_types}.

Position-sensitive detectors with continuous scintillator require a reconstruction algorithm: signals from photomultipliers are analyzed to calculate the coordinates of the scintillation flash. It is solved in several ways. The first approach is to use Anger's method \cite{ANGER1, ANGER2, ANGER3}. The coordinates of the flash are calculated as "center of mass", where the energy deposited in each photomultiplier is considered as "mass". This method is linear, which restricts its precision due to systematic errors and the significant effect of the so-called "field of view" - the area where linear approximation is reliable is smaller than the scintillator size \cite{overwiew_anger}. A more complex approach involves Monte Carlo modeling \cite{MONTE-CARLO1, MONTE-CARLO2} with some experimental calibration. First, an array of intensities for all possible points, where flash can happen, is modeled. Second, an inverse problem is solved. Inverse algorithms minimize the difference between intensity achieved in the experiment and the set of intensities from the modeled array under some metric to find the closest result in it.

Detectors with continious scintillators achieve higher precision when the scintillator faces are fully absorbing - adding reflectors can reduce coordinate resolution and field of view \cite{overwiew_anger}. However, it increases the energy resolution due to less light loss. The development of an algorithm capable of restoring the coordinate of a scintillation flash in a detector with a continuous scintillator with reflecting faces can lead to an increase in the energy resolution of such detectors while maintaining the same coordinate resolution. 

The problem of the coordinate reconstruction can be formalized. If a uniform flash with overall energy $I$ happens at the point $(x_0,y_0,z_0)$, there is a mapping \\$\vec{F}(x_0,y_0,z_0,I)$ that translates $x_0,y_0,z_0,I$ into a set of intensities $\vec{W}=\vec{F}(x_0,y_0,z_0,I)$. The component $W_i$ is the energy $i$-th photomultiplier gets. The inverse problem must be solved: an inverse mapping $\{\tilde{x}_0,\tilde{y}_0,\tilde{z}_0,I\}=\vec{F}^{-1}(\vec{W})$ must be found. It can be solved by minimising deviation under the norm  \begin{align}\|\vec{F}(\tilde{x}_0,\tilde{y}_0,\tilde{z}_0,I)-\vec{W}\|\label{norm}.\end{align}
Another way to solve the inverse problem is the Maximum likelihood estimation (MLE) method \cite{MLE}. The idea is a little bit more general: find the extremum of functional $M(\vec{F}(\tilde{x}_0,\tilde{y}_0,\tilde{z}_0,I),\vec{W})$, which depends on both vectors, while in Eq. (\ref{norm}) norm depends on difference $\vec{F}(\tilde{x}_0,\tilde{y}_0,\tilde{z}_0,I)-\vec{W}$. Conditional probability is a such type of functional. Thus, the solved maximization task is:
\begin{equation}
    (\tilde{x}_0,\tilde{y}_0,\tilde{z}_0) = Argmax( P(\vec{W}|XYZ)),
\end{equation}
here $P(\vec(W)|XYZ)$ is a probability of observing $\vec{W}$ photomultiplier signals with coordinates of the flash $(x,y,z)$. This could be described by a Poisson distribution \cite{MLE}.

The mentioned methods for scintillation flash coordinates reconstruction rely on approximate solutions for $\vec{F()}$. 
To push the coordinate reconstruction algorithms further using any of these methods, we should first solve the forward model $\vec{F()}$ precisely.


In general, this problem is difficult to solve because each ray path must be taken into account.
However, there are some detector setups that allow for analytical estimations and analysis.
Detectors with continuous scintillators are usually designed as rectangular cuboid scintillators with reflective faces.
The lower surface is covered with a matrix of silicon photomultipliers \cite{SiPM}, which are usually square. 
Reflectors can be either Lambertian or specular. 
Lambertian reflection obeys Lambert's cosine law \cite{Lambertian_Specular}.
For PET systems constructed from plastic scintillator strips \cite{moskal2016time} and for scintillation frustums with specular reflectors \cite{Specular_analytical} analytical estimations based on multiple ray reflections analysis to predict the outcome of the photomultipliers can be made.
To consider the Lambertian reflection, an iterative procedure can be involved \cite{evans2006monte} and analytical estimations can be made \cite{lambertian_analysis}.

In this article, we derive approximate analytical solutions for calculating the intensity outcome from the matrix of photomultipliers embedded in the scintillator bottom face (i.e., mapping $\vec{F}(x_0,y_0,z_0, I)$), while all other faces are considered to be specular reflectors with reflective indexes $\sim 1$ or $0$. 
The solutions are based on the kaleidoscopic reflection of the bottom face of the scintillator (further, we call this approach kaleidoscopic ray-tracing).
Numerically, this approach consists of the calculation of several infinite (or finite) sums and requires much fewer computations than Monte Carlo modeling, and can be implemented on a time scale of $0.1$ second, providing results similar to Monte Carlo modeling.
Further, using the derived analytical solutions, we reconstruct the coordinates of the scintillation flash.
The coordinate reconstruction using the developed approach shows similar spatial resolution as the coordinate reconstruction using Monte Carlo simulation.
Numerical implementations of the developed method assume errors, which are also estimated.
The analytical solutions are generalized to the cases where internal scattering, SiPM matrix geometric factor, and non-uniformity of the flash are important.

This paper is structured as follows.
In Section \ref{KRT_sec}, we explain the basic ideas behind the suggested kaleidoscopic ray-tracing method and get explicit equations for the intensities deposited in photomultipliers in general cases and for important particular cases.
In Section \ref{KRT_restr}, we address important aspects that can affect the implementation of the kaleidoscopic ray-tracing method in real life.
Namely, non-uniformity of light generation in a scintillation flash, scattering of the generated light, influence of the SiPM matrix geometric factor, and geometrical restriction of the kaleidoscopic ray-tracing method it involves by itself.
In Section \ref{Num_imp}, we discuss possible numerical implementations of the kaleidoscopic ray-tracing method and the errors these implementations involve.
In Section \ref{Simul_G4}, we discuss the usage of the GEANT4 simulation toolkit for Monte Carlo simulation of scintillations in the considered detectors to check the results provided by the kaleidoscopic ray-tracing method via direct Monte Carlo simulation of scintillation flashes.
In Section \ref{Comp_sec}, we compare the resolution that can be achieved when the kaleidoscopic ray-tracing method is used as a reference signal to initiate the flash coordinates reconstruction procedure to the resolution that can be achieved when Monte Carlo simulation is used instead.
In Section \ref{Conc_sec}, we summarize achieved results, draw conclusions, and discuss possible applications of the developed method.

\section{Kaleidoscopic ray-tracing method} \label{KRT_sec}
In this section, we explain the basic ideas behind the suggested kaleidoscopic ray-tracing method (KRT).
This method allows for a straightforward consideration of a tricky situation when multiple reflections of a significant number of rays produced by a scintillation flash happen.

\begin{figure}
    \centering
    \includegraphics[width=0.8\linewidth]{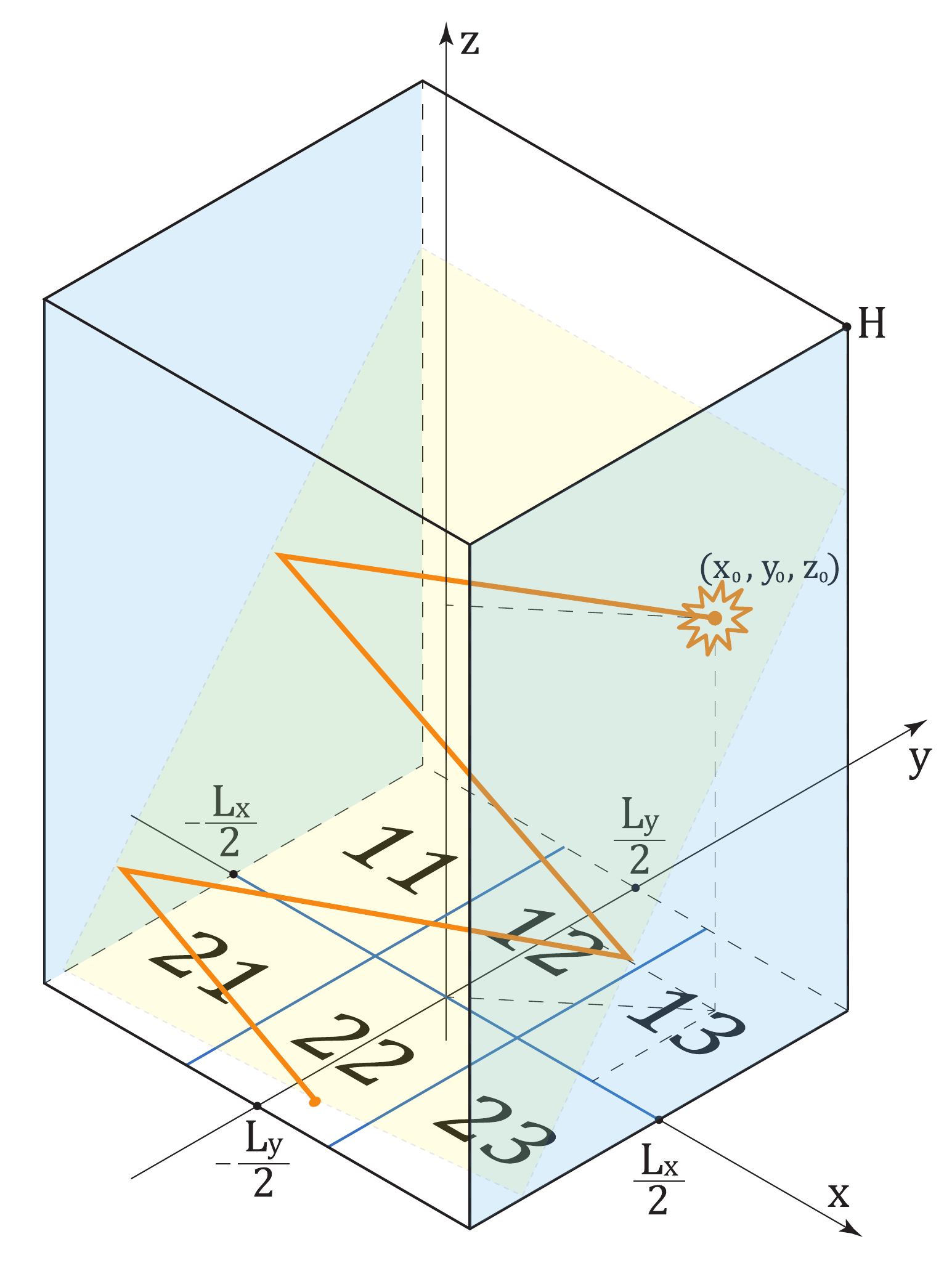}
    \caption{\label{The_Box} The scintillation crystal with reflective faces, $n=3$, $m=2$. The flash occurs at $(x_0,y_0,z_0)$ (orange source). A ray path is presented with the orange line. The yellow plane denotes the plane ray travels into. Photomultipliers are indexed as elements of a matrix.}
\end{figure}

Let us consider a rectangular cuboid scintillator with specular reflectors at the lateral and top faces.
The scintillator is placed as it is shown in Fig. \ref{The_Box}. The bottom of the scintillator is centered at the $(0,0,0)$ point and has $z=0$, the top has $z=H$. The bottom of the scintillator is split into $n \times m$  photomultipliers with the size $a\times b$, here $a=L_x/n$, $b=L_y/m$, here $L_x$ is the length of the scintillator and $L_y$ is the width (for simplicity and for clarity all the figures display the case $n=3$, $m=2$, the developed method is applicable for any case of $n \times m$). Here and further, if not specified, lengths and coordinates are given in arbitrary, but the same units (mm, cm, etc.).
The faces have corresponding reflective indexes $p_{xL}$ (the surface that is perpendicular to the $x$-axis and has $x<0$), $p_{xR}$ (the surface that is perpendicular to the $x$-axis and has $x>0$), $p_{yL}$ (the surface that is perpendicular to the $y$-axis and has $y<0$), $p_{yR}$ (the surface that is perpendicular to the $y$-axis and has $y>0$), $p_{zL}$ (bottom face of the scintillator), $p_{zR}$ (top face of the scintillator). 
The reflective index represents the ability of the surface to reflect electromagnetic energy. 
It is defined as the ratio of reflected ray intensity and incident ray intensity. In general case, reflective index depends on the angle of incidence \cite{born2013principles}. 
We consider all reflective surfaces, except the bottom face, to be black painted (reflective index is $0$) or to be made of metal with $\sqrt{\varepsilon}=n+i\varkappa$, $n\ll\varkappa$, here $\varepsilon$ is dielectric constant. For such materials reflective index is approximately constant and it is close to~$1$ for both polarisations.

\begin{figure}
    \centering
    \includegraphics[width=0.8\linewidth]{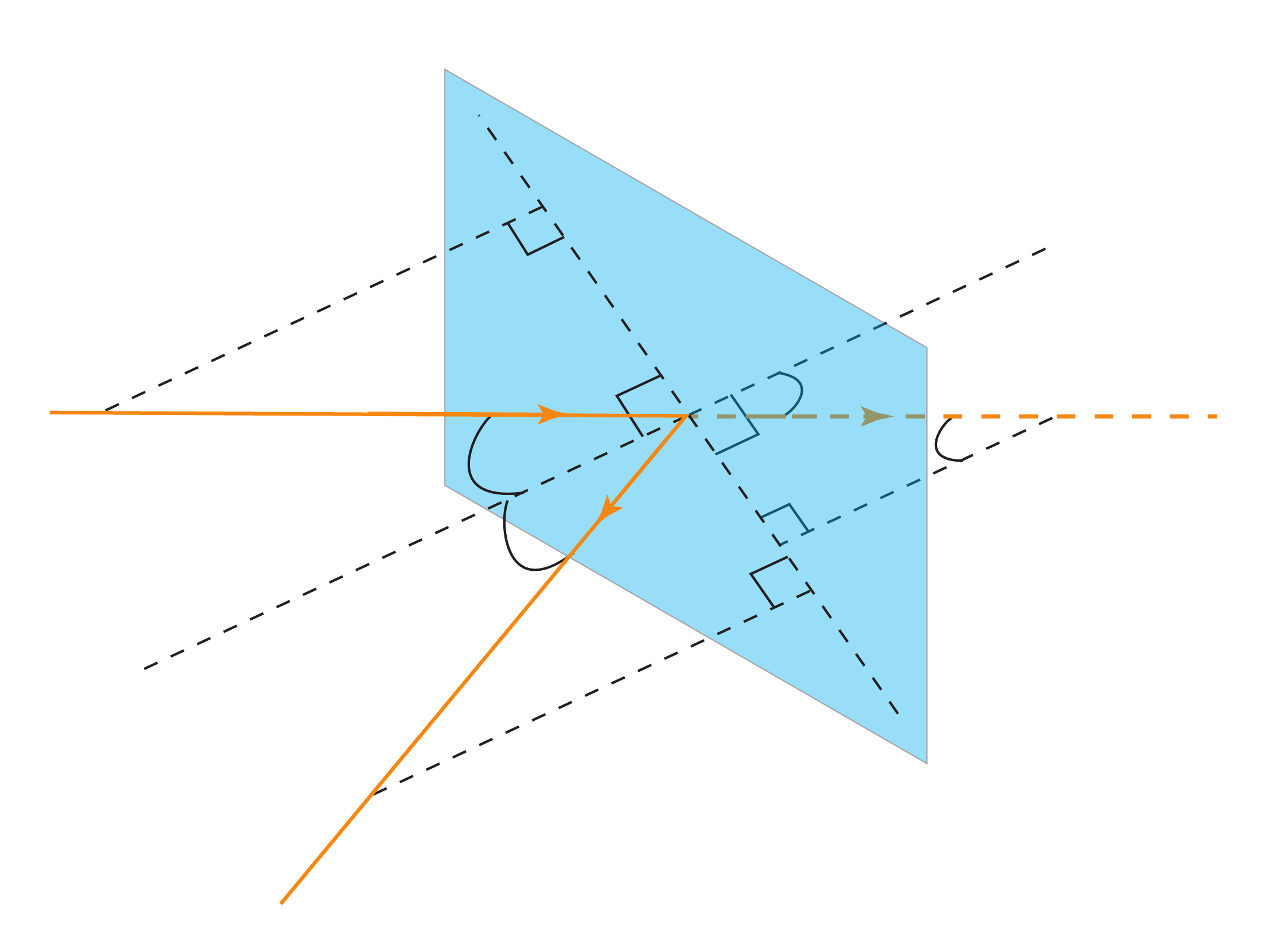}
    \caption{\label{The_reflect} An example of a ray, reflected by a lateral face of the scintillator (solid orange line) and its continuation over the reflective plane (dashed orange line).}
\end{figure}

A flash with energy $I$ that is uniformly distributed over the spherical angle occurs at $(x_0,y_0,z_0)$. Since we are only interested in the overall energy outcome and do not take into account the duration of the scintillation flash, we will not make any difference between energy and intensity.
To calculate the energy each of the photomultipliers receives, we kaleidoscopically reflect the bottom face of the scintillator over the wall faces as it is shown in Fig. \ref{Koleyido}. When a ray hits a wall face, the face reflects the ray at the same angle to the normal vector of the face (see Fig. \ref{The_reflect}). If now one reflects the internal space of the scintillator over the hit face, the ray will be straightened due to the laws of reflection as shown in Fig. \ref{The_reflect}. When this trick is applied to all the reflections from the scintillator wall faces, $xy$-plane becomes covered by scintillator bottom face reflections as shown in Fig. \ref{Koleyido}. Green lines denote reflections of the wall faces, blue lines depict reflections of the photomultipliers borders, black lines denote the wall faces of the scintillator. The solid orange line denotes the projection of the ray from Fig. \ref{The_Box} to the $xy$-plane. The dashed orange line denotes the projection of the straightened version of this ray. All green lines have their prescribed reflective indexes.
This procedure generates the same tessellation of $xy$-plane for different rays due to the cuboid shape of the scintillator.

Further, the photomultipliers are indexed with $i,j$, and reflections of scintillator internal space are indexed with $\alpha, \beta$. For instance: the real scintillator (black rectangle in Fig. \ref{Koleyido}) has $\alpha=0$, $\beta=0$, its reflection once over the wall with $x<0$ and twice over the wall with $y<0$ has $\alpha=-1$, $\beta=-2$, the reflection of the scintillator internal space once over the wall $y>0$ has $\alpha=0$, $\beta=1$, (see also Fig. \ref{Koleyido_the_box}).

After rays have been straightened, the wall faces do not affect the trajectories of rays. Then the intensity at the $ij$ photomultiplier is equal to the sum of intensities on all its reflections in the $xy$-plane. If the flash is uniform, the wall faces are absolutely reflective, and the top face is absolutely absorbing, one just should find the spherical angle occupied by reflections of the $ij$ photomultiplier from the point of view $(x_0,y_0,z_0)$ (see Fig. \ref{Koleyido_the_box}, yellow pyramid shows the spherical angle of $\alpha=-1$, $\beta=-1$ reflection of $i=1$, $j=2$ photomultiplier).

To compute the intensity at the $ij$ photomultiplier when the wall faces are reflective, one should sum the intensities at all reflections of the $ij$ photomultiplier, multiplying them by the reflective index prescribed to the green line, if the projection of the straightened trajectory of the ray to the $xy$-plane intersect this green line in Fig. \ref{Koleyido} and hits corresponding reflection of $ij$ photomultiplier. For instance: the intensity of the ray in Fig. \ref{Koleyido} should be multiplied by $p_{xL}^2p_{xR}$.

\subsection{Application of the KRT method to the case of fully transmitting bottom face}

A fully transmitting bottom face should be considered first to clearly explain the concept of KRT.
Here, fully transmitting means that the energy of all rays touching the bottom face of the scintillation crystal is transmitted to the photomultipliers (i.e. $p_{zL}=0$). 
Following the above, the intensity at the $ij$ photomultiplier should be calculated as

\begin{align}\label{I}
    \nonumber
     I_{ij}&=\sum_{\alpha=-\infty}^{+\infty}\sum_{\beta=-\infty}^{+\infty}p_{xL}^{n_{xL}}p_{xR}^{n_{xR}}p_{yL}^{n_{yL}}p_{yR}^{n_{yR}}\times \\ 
     &\times\left(\Delta I^{(x_0,y_0,z_0)}_{\alpha\beta,ij}+ p_{zR}\Delta I^{(x_0,y_0,2H-z_0)}_{\alpha\beta,ij}\right). 
\end{align}

Here $\Delta I^{(x_0,y_0,z_0)}_{\alpha\beta,ij}$ is the amount of energy deposited in the $\alpha\beta$ reflection of $ij$ photomultiplier from the flash that occurs at $(x_0,y_0,z_0)$, if the top face is fully absorbing. Since the top face is reflecting, one should take $\Delta I^{(x_0,y_0,2H-z_0)}_{\alpha\beta,ij}$ (the image source) under consideration. Its intensity should be multiplied by $p_{zR}$ -- the reflective index of the top face.

\begin{figure}
    \centering
    \includegraphics[width=1.02\linewidth]{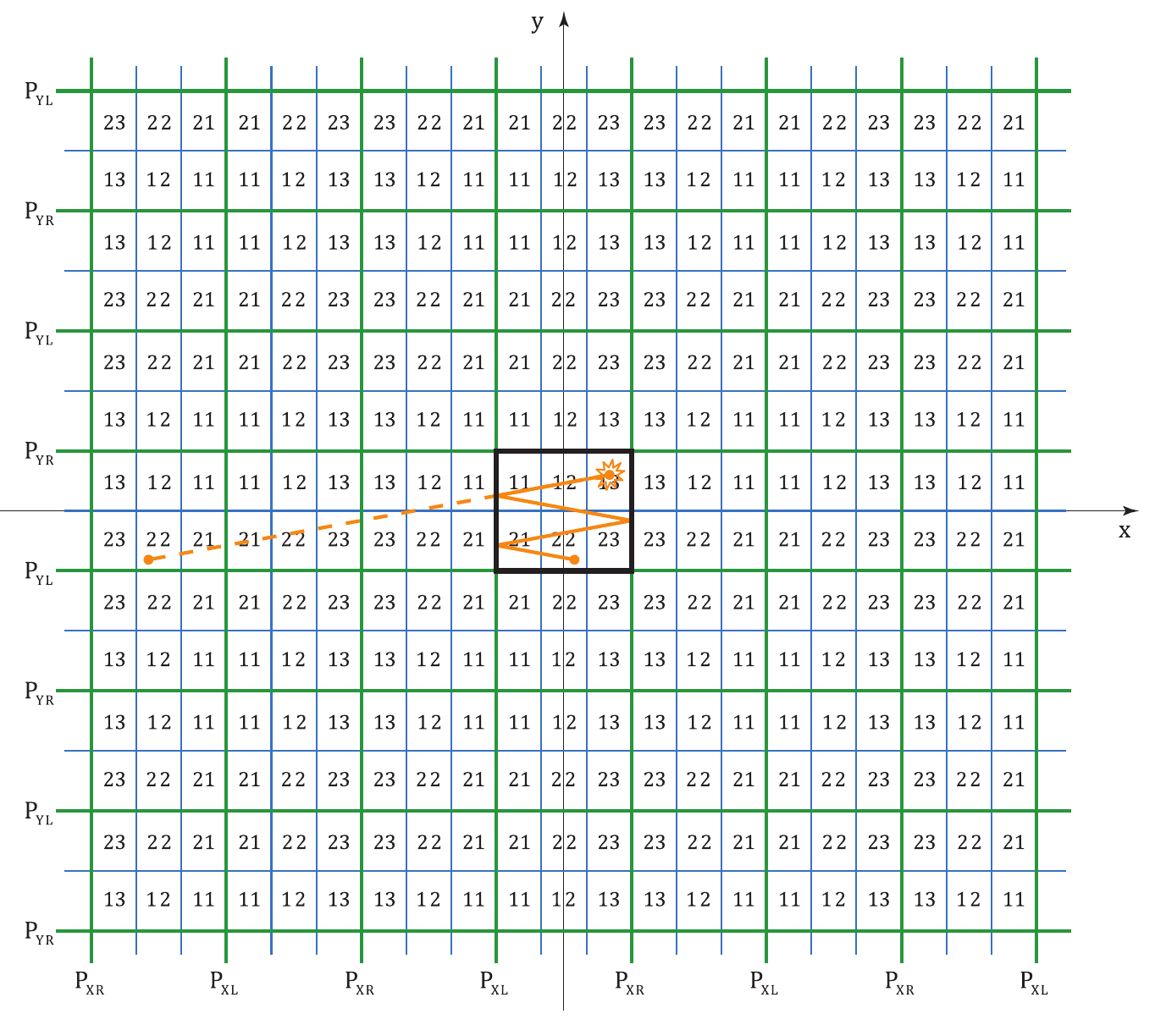}
    \caption{\label{Koleyido} Kaleidoscopic reflection of the scintillator bottom face (the bottom face of the scintillator is restricted with the black line) over the wall faces. Reflections of wall faces (green lines) have corresponding reflective indexes $p_{xL}$, $p_{xR}$, $p_{yL}$, $p_{yR}$. Axes coincide with corresponding axes from Fig. \ref{The_Box}. The solid orange line depicts the projection of the ray from Fig. \ref{The_Box} and the dashed orange line depicts its continuation over reflective faces.}
\end{figure}

There are three light beams: (i) the first one is formed by the rays moving downward (their contribution to the detected intensity is taken into account via $\Delta I^{(x_0,y_0,z_0)}_{\alpha\beta,ij}$ terms), (ii) the second beam is formed by the rays parallel to the bottom face (these rays are stuck in their plane and never reach the bottom face), (iii) the third beam is formed by the rays that move upward (their contribution to the detected intensity is taken into account via $\Delta I^{(x_0,y_0,2H-z_0)}_{\alpha\beta,ij}$ terms multiplied by $p_{zR}$ (this means that the additional source is placed as an imaginary source of the initial flash above the top face with initially reduced intensity by $p_{zR}$, because rays from the third beam touch the bottom face after being reflected by the top face). Degrees $n_{xL}$, $n_{xR}$, $n_{yL}$, $n_{yR}$ are the numbers of times the horizontal projection of a straightened ray intersect green lines in Fig. \ref{Koleyido} with the corresponding reflective index before it reaches the $\alpha\beta$ reflection of $ij$ photomultiplier. These numbers do not depend on the chosen ray for fixed $\alpha\beta$, $ij$.

\begin{figure}
    \centering
    \includegraphics[width=\linewidth]{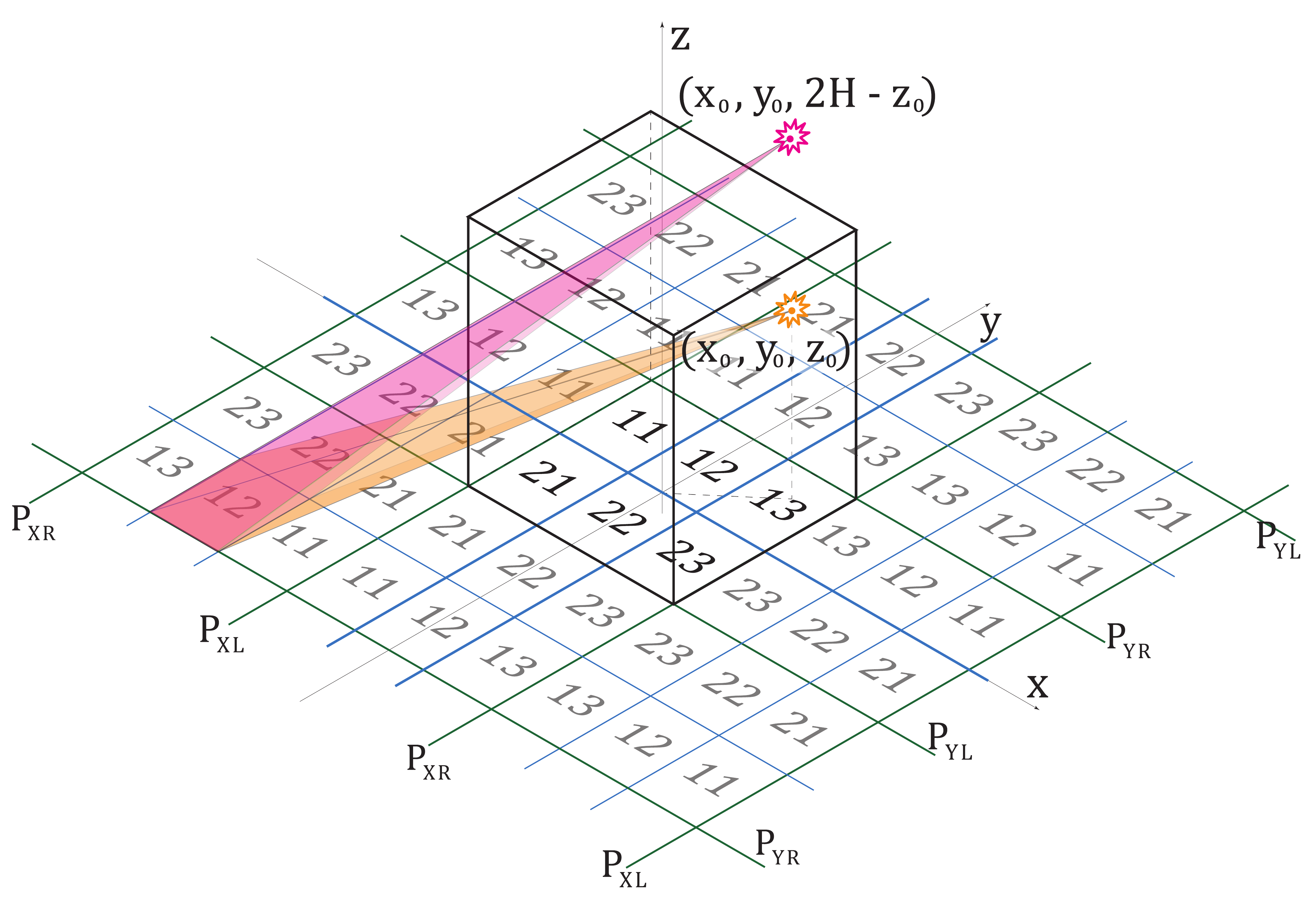}
    \caption{\label{Koleyido_the_box} Kaleidoscopic reflection of the scintillator bottom face over the wall faces (see Fig. \ref{Koleyido}) and spherical angle of the $\alpha=-1,\beta=-1$ reflection of the $i=1,j=2$ photomultiplier due to the flash (orange source) and its reflection over the top face (pink source).}
\end{figure}

To calculate the amount of energy per arbitrary rectangle area of length $a$ and width $b$ (further this rectangle area is denoted as $A$), it is necessary to calculate the integral of the Umov-Poynting vector component that is normal to the rectangle area $A$ over area $A$ \cite{LANDAU_FT, POYNTING}. For this purpose, the source with intensity $I$ at $\vec{r}_0=(x_0,y_0,z_0)$ is considered; rectangle $A$ lies in the $xy$-plane and its sides are parallel to $x$ and $y$-axis and its bottom left corner has $(x^{(L)},y^{(L)})$ coordinates. The mentioned integral can be written as:

\begin{equation}\label{I_rec}
    I_A=\Bigg|\int\limits_A (\vec{S},\vec{n}_A) dx dy\Bigg|=\Bigg|\int\limits_{x^{(L)}}^{x^{(L)}+a}dx\int\limits_{y^{(L)}}^{y^{(L)}+b}dy\frac{I}{4\pi R^2}\frac{z_0}{R}\Bigg|.
\end{equation}
Here \(\vec{n}_A\) is a normal vector of the rectangle area $A$, $\vec{R}=\vec{r}-\vec{r_0}$, $\vec{r}=(x,y,z)$ $R=|\vec{r}-\vec{r_0}|$, $\vec{S}=\frac{\vec{R}}{R}\frac{I}{4\pi R^2}$ - Umov-Pointing vector. In other words, this integral computes the spherical angle of the rectangle area $A$ from the point of view $(x_0,y_0,z_0)$. This integral is equal to (see Appendix for the derivation)

\begin{align}\label{I_A}
\nonumber
    &I_A=I\Bigg|\frac{\mathrm{sign}(y^{(L)}-y_0+b)}{4\pi}\Bigg(\rm{Tan^{-1}}\left(a,b\right)-\rm{Tan^{-1}}\left(0,b\right)\Bigg)\\
    &-\frac{\mathrm{sign}(y^{(L)}-y_0)}{4\pi}\Bigg(\rm{Tan^{-1}}\left(a,0\right)-\rm{Tan^{-1}}\left(0,0\right)\Bigg)\Bigg|.
\end{align}
\begin{align}
    &\rm{Tan^{-1}}(a,b)=\\ \nonumber
    &=\tan^{-1}\left(\frac{(x^{(L)}-x_0+a)|y^{(L)}-y_0+b|}{z_0\sqrt{(x^{(L)}-x_0+a)^2+(y^{(L)}-y_0+b)^2+z_0^2}}\right).
\end{align}

The degrees of reflective indexes can be calculated as:
\begin{equation}\label{xL}
    n_{xL}=\left\lfloor  \frac{\left|\alpha\right|+\frac{1-\mathrm{sign}(\alpha)}{2}}{2}  \right\rfloor,
    n_{xR}=\left\lfloor  \frac{\left|\alpha\right|+\frac{1+\mathrm{sign}(\alpha)}{2}}{2} \right\rfloor,
\end{equation}
\begin{equation}\label{yR}
    n_{yL}=\left\lfloor  \frac{\left|\beta\right|+\frac{1-\mathrm{sign}(\beta)}{2}}{2}  \right\rfloor,
    n_{yR}=\left\lfloor  \frac{\left|\beta\right|+\frac{1+\mathrm{sign}(\beta)}{2}}{2}  \right\rfloor. 
\end{equation}
Here $\lfloor \cdot\rfloor$ denotes the floor function: it takes a real number as input and gives the greatest integer less than or equal to the initial number.
Eq. (\ref{I_A}) and Eqs. (\ref{xL})-(\ref{yR}) give explicit answer being substituted in Eq. (\ref{I}). One should only define $x^{(L)}$ and $y^{(L)}$ for each of the reflections of the photomultipliers. We start with defining the centers of the photomultipliers $x^{(c)}_{00,ij}$ $y^{(c)}_{00,ij}$ as coordinates of points where diagonals of a photomultiplier intersect each other. Using properties of reflection it is easy to derive that

\begin{equation}\label{xc}
    x^{(L)}_{\alpha \beta,ij}=x^{(c)}_{00,ij}\cdot(-1)^{|\alpha|\text{\ } \mathrm{mod}\text{\ } 2} +\alpha L_x -a/2,
\end{equation}

\begin{equation}\label{yLL}
    y^{(L)}_{\alpha \beta,ij}=y^{(c)}_{00,ij}\cdot(-1)^{|\beta|\text{\ } \mathrm{mod}\text{\ } 2} +\beta L_y -b/2.
\end{equation}
Here $x^{(L)}_{\alpha \beta,ij}$ and $y^{(L)}_{\alpha \beta,ij}$ are coordinates of bottom left corner of the $\alpha\beta$ reflection of the $ij$ photomultiplier.

Eqs. (\ref{xL})-(\ref{yLL}) are just a convenient form for the number of reflections to be computed. Any other functions that give the same results out of $\alpha,\beta$ are suitable.

The algorithm of the calculation is as follows. One should use Eqs. (\ref{xc})-(\ref{yLL}) to calculate down left corners 
 of the $\alpha\beta$ reflection of the $ij$ detector. This results should be used in the Eq. (\ref{I_A}) that calculates $I_{\alpha\beta,ij}^{(x_0,y_0,z_0)}$. Substituting this results in Eq. (\ref{I}) together with Eqs. (\ref{xL})-(\ref{yR}) that define the powers of the reflective indexes one gets the $I_{ij}$ intensity (i.e. intensity deposited in $ij$ photomultiplier). 
 All operations before the summation over $ij$ can be done in parallel using vector calculations.



\subsection{Application of the KRT method to the case of a reflecting bottom face}\label{RBF}
In this section, we consider the case of a non-zero reflective index of the scintillation crystal bottom face.
We get some explicit equations on the intensities deposited in the photomultipliers attached to the bottom face of the scintillation crystal using the KRT method.

Usually, the scintillation crystal with refractive index $n_c$ is connected to the matrix of photomultipliers by a thin layer of a dielectric medium with refractive index $n_m$. The reflective index of the border crystal-medium can be expressed via Fresnel equations \cite{born2013principles}:
\begin{equation}
    r_s=\Bigg|\frac{\cos{\phi}-\sqrt{n^2-\sin^2{\phi}}}{\cos{\phi}+\sqrt{n^2-\sin^2{\phi}}}\Bigg|^2,
\end{equation}
\begin{equation}
    r_p=\Bigg|\frac{n^2\cos{\phi}-\sqrt{n^2-\sin^2{\phi}}}{n^2\cos{\phi}+\sqrt{n^2-\sin^2{\phi}}}\Bigg|^2,
\end{equation}
\begin{equation}
    r=(r_s+r_p)/2.
\end{equation}
Here $n=n_m/n_c$, $\phi$ is the angle of incidence. Reflective index $r_s$ is for s-polarized wave, reflective index $r_p$ is for p-polarized wave, and reflective index $r$ is for naturally polarized light. In this case, the reflective index of the bottom face depends on the incidence angle, so Eq. (\ref{I}) and Eq. (\ref{I_rec}) should be modified. One should replicate kaleidoscopic surface from Fig. \ref{Koleyido_the_box} at heights $-2H$, $-4H$, $-6H$, $\dots$. This trick straightens the rays that are reflected from the bottom face, and the previously considered procedure can be implemented again. Finally,

\begin{align}\label{I_nzL}
I_{ij}&=\sum_{n_{zL}=0}^{+\infty}p_{zR}^{n_{zL}}\sum_{\alpha=-\infty}^{+\infty}\sum_{\beta=-\infty}^{+\infty}p_{xL}^{n_{xL}}p_{xR}^{n_{xR}}p_{yL}^{n_{yL}}p_{yR}^{n_{yR}}\times \\ \nonumber
     &\times\left(\Delta I^{(x_0,y_0,z_0)}_{\alpha\beta,ij,n_{zL}}+ p_{zR}\Delta I^{(x_0,y_0,2H-z_0)}_{\alpha\beta,ij,n_{zL}}\right), 
\end{align}
\begin{align}\label{I_rec_nzL} &\Delta I^{(x_0,y_0,z_0)}_{\alpha\beta,ij,n_{zL}}=\\ \nonumber
&\int\limits_{A_{\alpha\beta,ij,n_{zL}}}\frac{Idxdy}{4\pi R^2_{n_{zL},z_0}}\frac{z_0+2Hn_{zL}}{R_{n_{zL},z_0}} \frac{(1-r_p)r^{n_{zL}}_p+(1-r_s)r^{n_{zL}}_s}{2}.
\end{align}
Here $R_{n_{zL},z_0}=\sqrt{(x-x_0)^2+(y-y_0)^2+(z_0+2Hn_{zL})^2)}$, $\Delta I^{(x_0,y_0,z_0)}_{\alpha\beta,ij,n_{zL}}$ denotes amount of energy that is deposited in $\alpha\beta$ reflection of $ij$ photomultiplier in $n_{zL}$ copy of the $xy$-plane from the flash that occurs at $(x_0,y_0,z_0)$, if there is no top face. The $A_{\alpha\beta,ij,n_{zL}}$ is the integration region. The intensity of the imaginary source is also considered.

\begin{figure}
    \centering
    \includegraphics[width=0.9\linewidth]{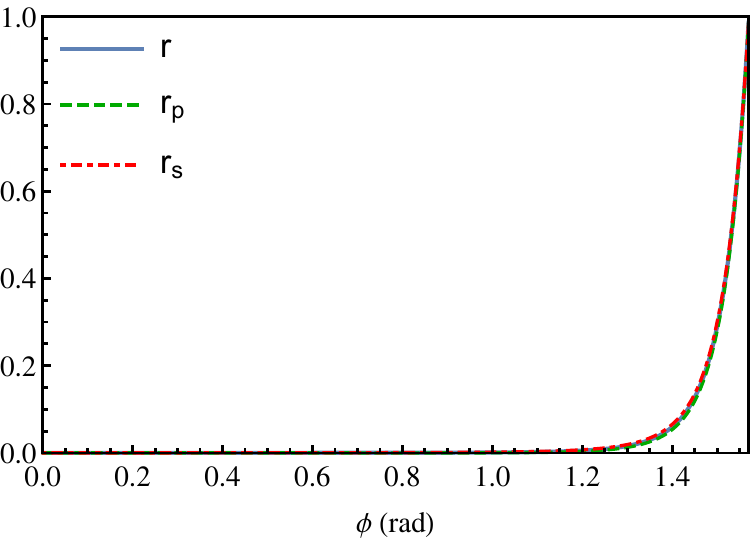}
    \caption{\label{Ref_195} The dependence of $r$ (blue solid line), $r_p$ (green dashed line), $r_s$ (red dot-dashed line) on the incidence angle $\phi$ in radians $n=1.95/1.9$.}
\end{figure}

If $n>1$, then the effect of total internal reflection does not exist.
If also $n\sim 1$, then the dependencies of $r_s$, $r_p$, $r$ on the angle of incidence are rather simple. For GAGG: Ce $n_c=1.9$ near the $540$ nm wavelength \cite{kozlova2018optical}. In Fig. \ref{Ref_195} the dependencies of $r$, $r_p$, $r_s$ on the incidence angle for $n=1.95/1.9$ are presented. Reflective indexes are approximately $0$ for incidence angles less than $85^\circ$ and sharply grow to $1$ while the incidence angle grows from $85^\circ$ to $90^\circ$ (further the angle region of the reflective indexes sharp growth is called transition range). Thus, the medium that satisfies both conditions $n_m>n_c$ and $n_m\approx n_c$ can be treated as approximately fully transmitting. For the case of approximately fully transmitting medium the Eq. (\ref{I}) can be applied. The error of this method can be estimated as the ratio of the spherical angle of the transition range to $2\pi$: $\sigma=\Omega_{TR}/2\pi=\cos{\theta}\approx 0.1=10\%$, here $\theta=84^\circ$: the angle greater than that the reflective index is sufficiently greater than $0$.

\begin{figure}
    \centering
    \includegraphics[width=0.9\linewidth]{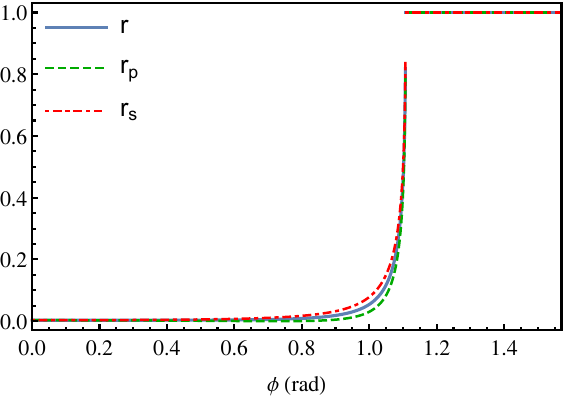}
    \caption{\label{Ref_150} The dependence of $r$ (blue solid line), $r_p$ (green dashed line), $r_s$ (red dot-dashed line) on the incidence angle $\phi$ in radians for $n=1.7/1.9$.}
\end{figure}

\begin{figure}
    \centering
    \includegraphics[width=\linewidth]{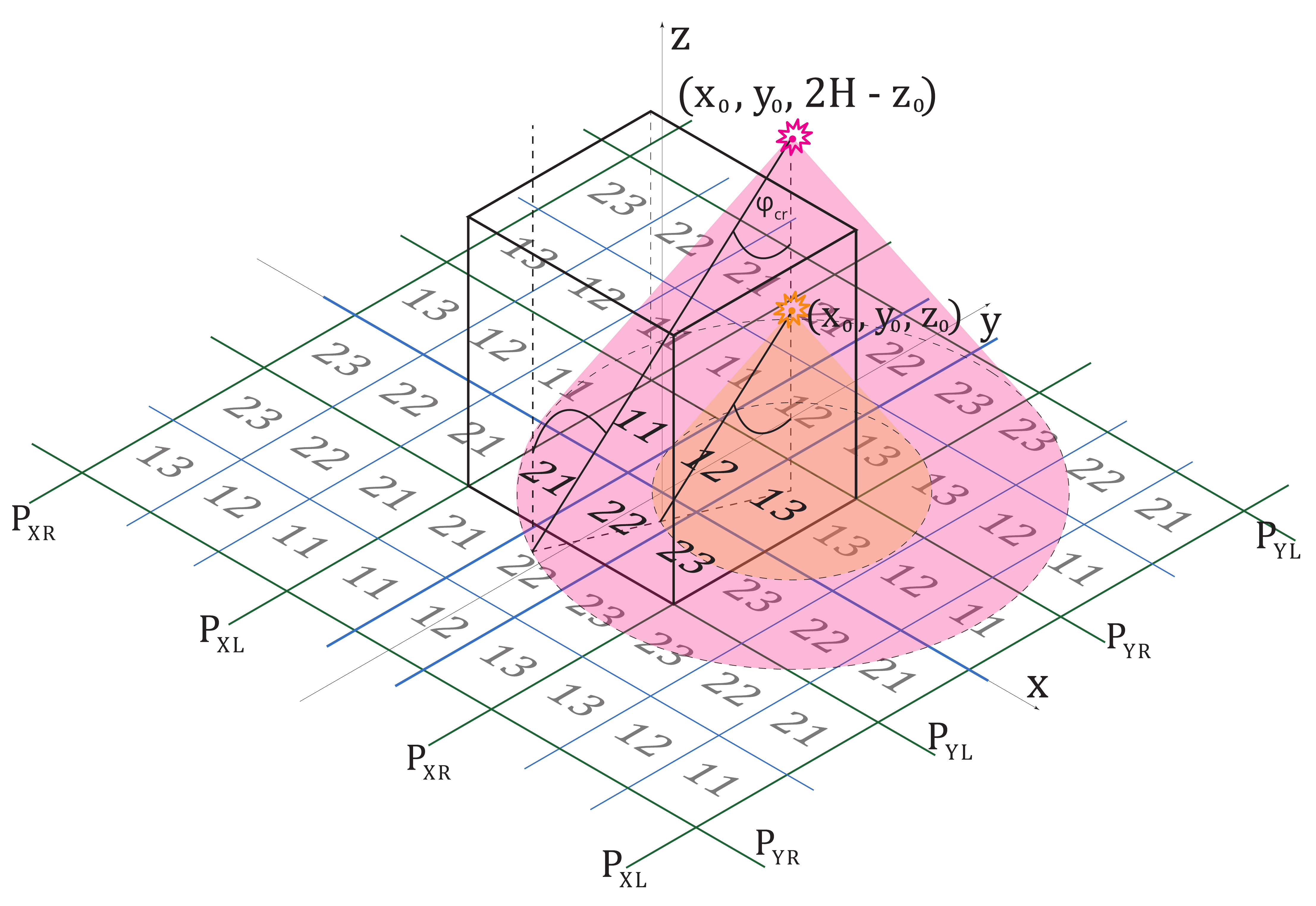}
    \caption{\label{Tot_ref} The critical cones of the real (orange cone) and imaginary (pink cone) flashes.}
\end{figure}

If $n<1$, total internal reflection occurs at the critical angle $\sin \phi_{cr}=n_m/n_c$. The dependence of $r$, $r_p$, $r_s$ on incidence angle is presented in Fig. \ref{Ref_150} for the case $n=1.7/1.9$. The reflective indexes are $0$ for incidence angle less than $60^\circ$, and sharply grows from $0$ to $1$, while incidence angle grows from $60^\circ$ to $\phi_{cr}\approx 64^\circ$, and are $1$ for the angles greater than $\phi_{cr}$. Thus, the rays that can be absorbed by the photomultipliers form a cone (further critical cone) with apex angle $2\phi_{cr}$ (see Fig. \ref{Tot_ref}). 

The bottom face in this case can be considered approximately fully transmitting but only in the overlap area between the critical cone and $xy$-plane. The Eq. (\ref{I}) can be applied here, but $\Delta I$ should be computed over the area of the critical cone and $xy$-plane overlap. 
The error of this method can be estimated as the ratio of the spherical angle of the transition range to the spherical angle of the critical cone: $\sigma=\Omega_{TR}/\Omega_{cr}=1-(1-\cos{60^\circ})/(1-\cos{64^\circ})\approx 0.1=10\%$. 
It should be mentioned that the case $n_c>n_m$ with condition $n_c\approx n_m$ can be considered as approximately fully transmitting case, with $\phi_{cr}\approx 90^\circ$.
Also, the following approximate analytical approach is applicable here:
\begin{align}\label{I_tot}
    \nonumber
     I_{ij}&=\sum_{\alpha=-\infty}^{+\infty}\sum_{\beta=-\infty}^{+\infty}p_{xL}^{n_{xL}}p_{xR}^{n_{xR}}p_{yL}^{n_{yL}}p_{yR}^{n_{yR}}\times \\ 
     &\times\left(\Delta \tilde{I}^{(x_0,y_0,z_0)}_{\alpha\beta,ij}+ p_{zR}\Delta \tilde{I}^{(x_0,y_0,2H-z_0)}_{\alpha\beta,ij}\right). 
\end{align}
Here $\Tilde{I}^{(x_0,y_0,z_0)}_{\alpha\beta,ij}=I^{(x_0,y_0,z_0)}_{\alpha\beta,ij}$, if $\alpha\beta,\ij$ reflection lies inside the circle, which is formed as the intersection of the critical cone and $xy$-plane; if $\alpha\beta,\ij$ reflection lies inside the circle partially
$\Tilde{I}^{(x_0,y_0,z_0)}_{\alpha\beta,ij}=I^{(x_0,y_0,z_0)}_{\alpha\beta,ij}S^{(x_0,y_0,z_0)}_{\alpha\beta,ij}/(ab)$, $S^{(x_0,y_0,z_0)}_{\alpha\beta,ij}$ is the area of overlapping between the circle and $\alpha\beta,\ij$ reflection.

Another option is to use an analytical solution that accounts for the spherical angles of the critical cone and a photomultiplier possible overlapping forms. 
These forms can be divided into rectangles and forms like the blue form in Fig. \ref{B_area}. 
The spherical angle of a form is equal to the sum of spherical angles of the rectangles and the forms like the blue form from Fig. \ref{B_area}. 
Intensity translated inside the spherical angle of the rectangle is considered in Eq. (\ref{I_A}). 
Intensity translated inside the spherical angle of the blue form from Fig. \ref{B_area} is equal to (see Appendix): 
\begin{figure}
    \centering
    \includegraphics[width=0.75\linewidth]{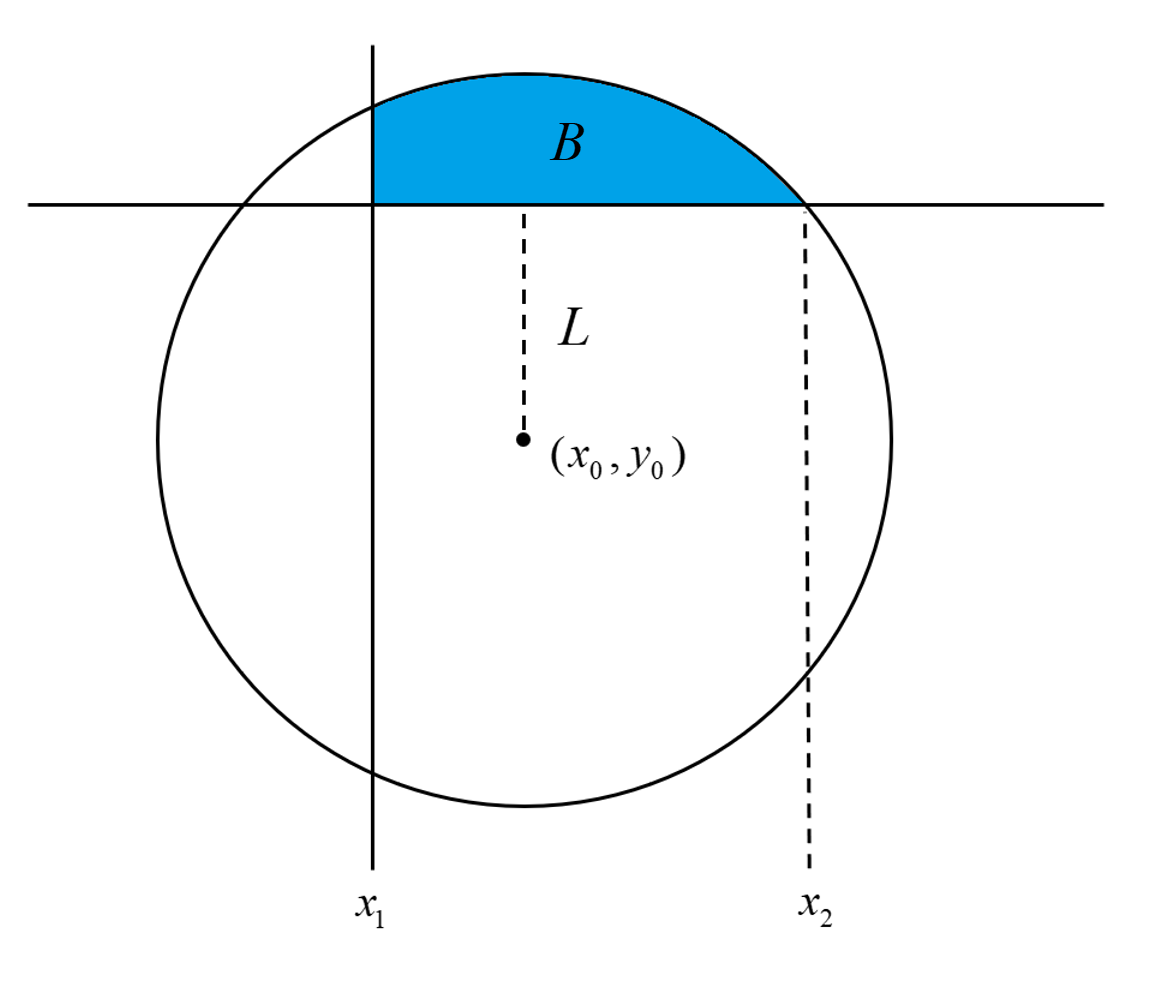}
    \caption{\label{B_area} Possible case of a critical cone bottom circle and
    photomultipliers overlapping}
\end{figure}

\begin{gather}\label{Dug}        
I_B=\iint_{B}\frac{Iz_0dxdy}{4\pi((x-x_0)^2+(y-y_0)^2+z_0^2)^{3/2}}\\ \nonumber
=\frac{Iz_0}{4\pi}(F(x_2-x_0,R_{cr},z_0,L)-F(x_1-x_0,R_{cr},z_0,L)).
\end{gather}
Here the function $F(u,R_{cr},z_0,L)$ is equal to
\begin{gather}\label{F_B}
F(u,R_{cr},z_0,L)=\\ \nonumber
=\frac{1}{\sqrt{R_{cr}^2+z_0^2}}\Bigg(\frac{\sqrt{R_{cr}^2+z_0^2}} {z_0}\mathrm{tan^{-1}}\left(\frac{u\sqrt{R_{cr}^2+z_0^2}}{z_0\sqrt{R_{cr}^2-u^2}}\right)-\\ \nonumber
-\mathrm{tan^{-1}}\left(\frac{u}{\sqrt{R_{cr}^2-u^2}}\right)\Bigg)-\\ \nonumber
-\frac{1}{z_0}\tan^{-1}\left(\frac{uL}{z_0\sqrt{L^2+z_0^2+u^2}}\right),  
\end{gather}
the $R_{cr}=z_0 \tan\phi_{cr}$ is the radius of the critical cone bottom circle, $x_{1,2}$ are $x$ coordinates where critical cone bottom circle intersects the photomultiplier sides, $L$ is the distance from $(x_0,y_0)$ point and blue area $B$ in Fig. \ref{B_area}.

Note that, due to the finite width of the connecting medium, an additional error appears in this approach.
Indeed, rays that travel near the lateral surface of the critical cone are almost parallel to the matrix of photomultipliers after the refraction at the crystal-connecting medium border.
Thus, these rays do not obey the geometry represented in Fig. \ref{Tot_ref}, as they can avoid being detected by a photomultiplier they've been initially pointed to.

This error can be estimated as follows.
Let $d$ be the width of the connecting medium here.
The relative amount of such rays can be estimated as $d(2L_x/n+2L_y/m)/(L_x/n+L_y/m)^2=2d/(L_x/n+L_y/m)$.
Usually $d\approx 0.1$ mm and $L_x/n\approx L_y/m\approx 5$ mm.
Thus, the additional error is of the order of $2\%$.

\section{Several important cases and restrictions of the KRT method}\label{KRT_restr}
In this section, we address several important details that can affect the implementation of the KRT method in real life and the restrictions of the KRT method that it involves by itself.

\subsection{Non-uniform flash}




If the intensity of the flash has non-uniform angle distribution, Eq. (\ref{I}) stays true. 
Formulas for degrees of reflective indexes also stay true.
However, Eqs. (\ref{I_rec})-(\ref{I_A}) should be modified.
Let $P(\theta,\phi)$ be the distribution of the intensity over the sphere with radius one.
Thus, the next formula for intensity can be used instead of Eq. (\ref{I_rec}):
\begin{align}\label{I_rec_ugl}
    &I_A=\Bigg|\int\limits_A (\vec{S}(l),\vec{n})dx dy\Bigg|=\\ \nonumber
    &=\Bigg| \int\limits_{x^{(L)}}^{x^{(L)}+a}dx\int\limits_{y^{(L)}}^{y^{(L)}+b}dy \frac{P(\theta,\phi)}{ R^2}I\frac{z_0}{R}\Bigg|.
\end{align}

Non-uniformity of the flash can be taken into account by applying Eq. (\ref{I_rec_ugl}) instead of Eq. (\ref{I_rec}) in Eq. (\ref{I}).

\subsection{SiPM matrix geometric factor}

Usually, photomultipliers are separated one from another by non-zero width bands \cite{ANGER3, SiPM1}. 
Thus the photomultipliers cover the bottom face only partially. 
Let the width of the band between photomultipliers be $2\kappa$. 
The existence
of bands reduces the amount of energy each of photomultipliers
get by reducing the spherical angles of the
photomultipliers and the spherical angles of their reflections,
compared to the case of zero width bands.
The impact of this effect can be taken into account by inserting frames with width $\kappa$ into all reflections of each photomultiplier. This way $I_A$ should be calculated as: 

\begin{equation}
    I_A=\Bigg|\int\limits_{x^{(L)}+\kappa}^{x^{(L)}+a-\kappa}dx\int\limits_{y^{(L)}+\kappa}^{y^{(L)}+b-\kappa}dy\frac{I}{4\pi R^2}\frac{z_0}{R}\Bigg|.
\end{equation}


\subsection{Contribution of the internal light scattering}

The contribution of the internal light scattering can be taken into account by means of Bouguer–Lambert-Beer law \cite{mayerhofer2020bouguer}.
According to it the intensity of a monochromatic wave exponentially decreases with inelastic mean free path $k_\lambda$ while the wave travels through a medium (here $\lambda$ is the wave length).
In the context of the addressed problem spectrum-angle distribution is unknown, so an approximate solution might be achieved for the case when the mean free path does not depend on the wave length $k_\lambda=k$.
First, the Eq. (\ref{I_rec}) needs to be rewritten for the case of the scattering medium.
\begin{equation}\label{I_LS}
    I_A=\Bigg|\int\limits_{x^{(L)}}^{x^{(L)}+a}dx\int\limits_{y^{(L)}}^{y^{(L)}+b}dy\frac{I}{4\pi R^2}\frac{z_0}{R}e^{-R/k}\Bigg|.
\end{equation}
Here the contribution of scattered light to the intensity outcome is neglected. The exponent in the integral may be approximately considered by multiplying intensities deposited in $\alpha\beta$ reflection of $ij$ photomultiplier in Eq. (\ref{I}) by corresponding exponents:
\begin{align}\label{I_k_exp}
     I_{ij}&=
     \sum_{\alpha=-\infty}^{+\infty}\sum_{\beta=-\infty}^{+\infty}p_{xL}^{n_{xL}}p_{xR}^{n_{xR}}p_{yL}^{n_{yL}}p_{yR}^{n_{yR}}\times \\ \nonumber
     &\times\Big(\Delta I^{(x_0,y_0,z_0)}_{\alpha\beta,ij} \exp{\left(-R_{\alpha\beta,ij}^{(x_0,y_0,z_0)}/k\right)}+\\\nonumber &+p_{zR}\Delta I^{(x_0,y_0,2H-z_0)}_{\alpha\beta,ij}\exp{\left(-R_{\alpha\beta,ij}^{(x_0,y_0,2H-z_0)}/k\right)}\Big).
\end{align}
Here $R_{\alpha\beta,ij}^{(x_0,y_0,z_0)}$ denotes the distance between the flash and the center of the $\alpha\beta$ reflection of the $ij$ photomultiplier (analogously for the imaginary flash source $R_{\alpha\beta,ij}^{(x_0,y_0,2H-z_0)}$):
\begin{equation}
R_{\alpha\beta,ij}^{(x_0,y_0,z_0)}=\sqrt{\big(x^{(c)}_{\alpha \beta,ij}-x_0\big)^2+\big(y^{(c)}_{\alpha \beta,ij}-y_0\big)^2+z_0^2}.
\end{equation}

In the particular case of GaGG:Ge scintillators with sizes $14\times14\times 2$ mm and $14\times14\times 7$ mm, the inelastic mean free path of the light is $64.5$ cm \cite{gagg_light_attenuation}. Thus, the light scattering contribution is sufficient only after 4-5-th reflection. According to Fig. \ref{Err}, after the 5-th reflection, $90\%$ of the flash energy is already collected even in the case of higher scintillator. Thus, the light scattering does not affect the energy resolution dramatically in the case of the considered scintillator. For higher scintillators Eqs. (\ref{I_LS})-(\ref{I_k_exp}) should be applied.

In general case, considering the contribution of the light scattering analytically in the most accurate way is an intricate and compound question. Thus, we leave a profound investigation of it for further research.

\subsection{Geometric limitations of the method}
To implement all the procedures that we stated above correctly, the shape of the scintillator should be one of kaleidoscopic forms. 
This provides us with an important property of kaleidoscopes: if we reflect the internal space $U$ of the kaleidoscope over its different walls several times, finally getting the reflection $U'$, the way how did $U$ was translated to $U'$ (over which wall it has been reflected and how many times) does not matter: the reflections of the photomultiplier $ij$ will coincide \cite{KALEIDOSCOPE}. 
This fact allows us to straighten reflected rays using the same tessellation of $xy$-plane for all rays.

\section{Numerical implementations of Eq. (\ref{I}) and Eq. (\ref{Dug})}\label{Num_imp}
In this section, we discuss numerical implementations of the KRT method and the errors these implementations involve.
An easy way to implement Eq. (\ref{I}) numerically is to consider $\alpha\in\{-N,\dots,N\}$, $\beta\in\{-M,\dots,M\}$: 

\begin{align}\label{I_N}
    \nonumber
     I_{ij}^{(N,M)}&=\sum_{\alpha=-N}^{N}\sum_{\beta=-M}^{M}p_{xL}^{n_{xL}}p_{xR}^{n_{xR}}p_{yL}^{n_{yL}}p_{yR}^{n_{yR}}\times \\ 
     &\times\left(\Delta I^{(x_0,y_0,z_0)}_{\alpha\beta,ij}+ p_{zR}\Delta I^{(x_0,y_0,2H-z_0)}_{\alpha\beta,ij}\right). 
\end{align}
Numbers $N$ and $M$ should be chosen in a way that limits the remaining terms in the sum at the needed level of error. Thus, the collected intensity is proportional to the spherical angle occupied by the rectangle $L_x(2N+1)\times L_y(2M+1)$ used in computations, and the error is proportional to the $2\pi$ minus mentioned spherical angle.

\begin{gather}\label{dela_I}
    \delta I=|I-I_{tot}^{(N,M)}|\le \\ \nonumber
    \le I\left(2\pi-\Delta\Omega^{(x_0,y_0,z_0)}+p_{zR}\left(2\pi-\Delta\Omega^{(x_0,y_0,2H-z_0)}\right)\right).
\end{gather}
Here $\Delta\Omega^{(x_0,y_0,z_0)}$ is the spherical angle occupied by the rectangle that is formed by considered reflections from the point of view $(x_0,y_0,z_0)$. The $I_{tot}^{(N,M)}=\sum_{ij}I_{ij}^{(N,M)}$ is total intensity on photomultipliers collected by mentioned rectangle. In Eq. (\ref{dela_I}) we considered absolutely reflective lateral faces to slow down the convergence of the series considering the hardest computational case. 

Let $r_m=\mathrm{min}((N+1/2)L_x,(M+1/2)L_y)$. If  $r_m\gg H$, then in Eq. (\ref{dela_I}) $x_0=0$ and $y_0=0$ can be considered. Finally, if $N\sim M$ the error can be evaluated as:

\begin{equation}
    \frac{\delta I}{I}\simeq\frac{2\pi r_m z_0}{4\pi r_m^2}+p_{zR}\frac{2\pi r_m (2H-z_0)}{4\pi r_m^2},
\end{equation}

\begin{equation}
    \frac{\delta I}{I}\simeq\frac{z_0}{2r_m}+p_{zR}\frac{  2H-z_0}{2 r_m}\sim\frac{1}{N}.
\end{equation}

\begin{figure}
    \centering
    \includegraphics[width=0.9\linewidth]{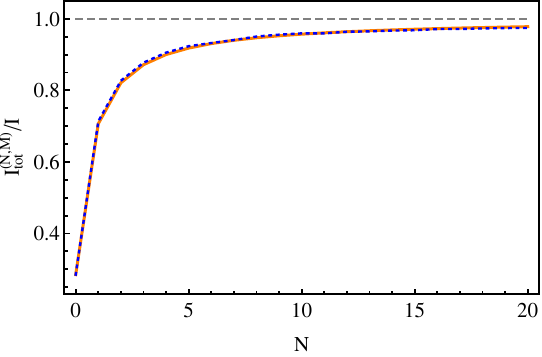}
    \caption{\label{Err} The dependence of $I_{tot}^{(N,M)}/I$ -- total normalized intensity deposited in photomultipliers via Eq. (\ref{I_N}) (orange line) and Monte Carlo simulation (blue dashed line) on $N$ in Eq.~(\ref{I_N}) ($M=N$ is used). The flash occurs in $x=5$, $y=-2$. Parameters are: $H=7$, $n=2$, $m=2$, $a=7$, $b=7$. All reflection indexes are equal to $1$, except $p_{zL}=0$.}
\end{figure}

The orange line in Fig. \ref{Err} depicts how total intensity on the photomultipliers $I_{tot}^{(N,M)}$ depends on the $N$ in Eq. (\ref{I_N}) in the case $N=M$. At $N=M=5$ more than $90\%$ of the energy is collected by the considered reflections and at $N=M=20$ the value of $\delta I/I\simeq0.03$.  

Numerical implementation of Eq. (\ref{Dug}) does not involve significant additional error by itself.
The error (compared to direct modeling) that appears, while Eq. (\ref{Dug}) is implemented, appears only due to the simplifications that have been made and discussed in Section \ref{RBF}.

\section{GEANT4 Monte Carlo simulation}\label{Simul_G4}
In this section, we discuss the Monte Carlo simulations we performed to check the compatibility of the achieved analytical results with data a close-to-real-life detector setup provides.
For that, we use GEANT4 \cite{GEANT4} simulation toolkit --- a widespread tool for the simulation of the scintillation physics. Optical physics is represented by a separate module, that implements different models. In our case, it was a UNIFIED model, adopted from DETECT \cite{DETECT}. To simulate a specular reflector, we used "dielectric\_dielectric" and "polishedfrontpainted" constants. In that case, only specular spike reflection or absorption was possible, without refraction.

\begin{figure}
    \centering
    \includegraphics[width=\linewidth]{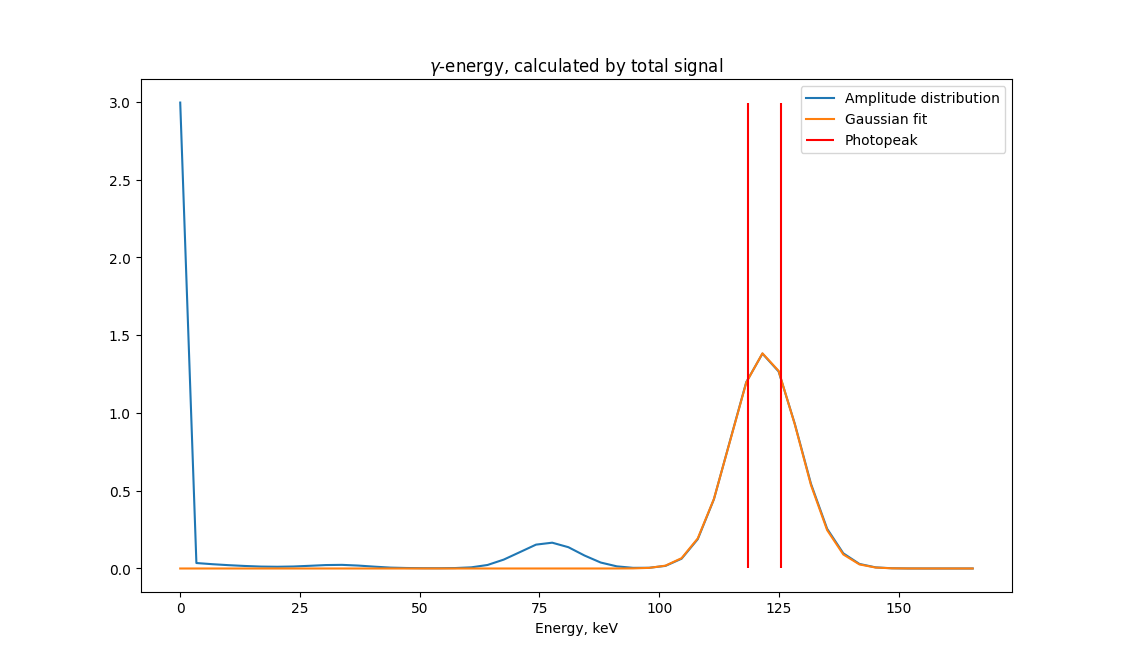}
    \caption{\label{122_spec} Simulated total signal calibrated for peak energy 122 keV.}
\end{figure}

The energy of the scintillation flash is defined by GEANT4 via the "SCINTILLATIONYIELD" constant, multiplied by gamma energy loss. This constant was set to 40 $keV^{-1}$ \cite{gagg_property}. Refractive indexes of the materials were not important due to "polishedfrontpainted" coatings of side and upper faces - only reflection or absorption was allowed, no refraction. The bottom surface was considered fully transmitting. Another important parameter, "RESOLUTIONSCALE", was set to 3.59 \cite{gagg_property} - it has the same meaning as the Fano factor for GAGG:Ce energy resolution - it defines statistical distribution for the number of produced scintillation photons. Scintillation photons had an average wavelength of 540 nm \cite{gagg_property}.

GEANT4 doesn't have a specific model for GAGG:Ce material, so it was constructed according to the chemical formula (Gd3 Al2 Ga3 O12) using the G4Material method - AddElement(). This material gives approximate radiation length and cross-sections of main processes: Compton scattering and photoeffect, in GAGG:Ce.

To test Eq. (\ref{I}) and Eq. (\ref{Dug}) and compare them with GEANT4 results, GAGG: Ce \cite{GAGG} scintillator $14\times14\times2$ mm size was placed in the center. This means that the scintillator had a 1 mm maximum coordinate and a -1 mm minimum coordinate along the $z$-axis. Its lower face was covered with 4 equal squares of size $7\times7$ mm, which detected photons with a probability of 0.3, simulating the silicon photomultiplier's \cite{SiPM} (usually used as detectors) efficiency. Then a 122 keV particle was launched into the scintillator along the $z$-axis.
Similarly, a scintillator of $14\times14\times7$ mm size was examined.

GEANT4 simulation provides binary data of the total energy outcome on each of the four SiPMs after the scintillation flash. Every $(x,y)$ point of the scintillator was struck by 122 keV gamma-quantum with the step of 0.5 mm along both axes for 10000 times. 

There is a significant number of Compton scatterings, as shown in Fig. \ref{122_spec}. Events with the energy outside of the range, bounded by red vertical lines, were excluded by analizing of the total signal on the matrix of attached photomultipliers. The intensity $I$ was calculated as the scintillation yield of GAGG:Ce multiplied by 122 keV energy, then multiplied by 0.3 to consider SiPM's efficiency.

\section{Comparison of the KRT method and Monte Carlo simulation}\label{Comp_sec}
In this section, we compare intensities deposited in photomultipliers calculated via the KRT method to intensities deposited in photomultipliers in Monte Carlo simulations.
After the comparison, we use both the KRT method and Monte Carlo simulation as reference signals to initiate flash coordinates reconstruction procedures and compare the precision achieved in both approaches.

\subsection{Simulation and flash coordinates reconstruction in the case of a fully transmitting bottom surface}\label{SFTBF}

Here we consider the case of a fully transmitting bottom surface of the scintillation crystal.
We show that both mentioned approaches give similar photomultipliers' outcomes. 
Also, in both approaches, we get similar results for the flash coordinate resolution: $1-1.8$ mm for the thin detector ($14\times14\times2$ mm) and $1.5-2.6$ mm for the thick detector ($14\times14\times7$ mm).
In both cases, intensity is collected by the matrix of four $7\times7$ mm photomultipliers.

First, we depict the total intensity deposited in photomultipliers in the Monte Carlo simulation for the crystal parameters analyzed in Fig. \ref{Err} by the dashed blue line in Fig. \ref{Err}.
The simulation results are similar to the results achieved using Eq.~(\ref{I}) in this case.

\begin{figure}[h]
    \centering
    \includegraphics[width=0.9\linewidth]{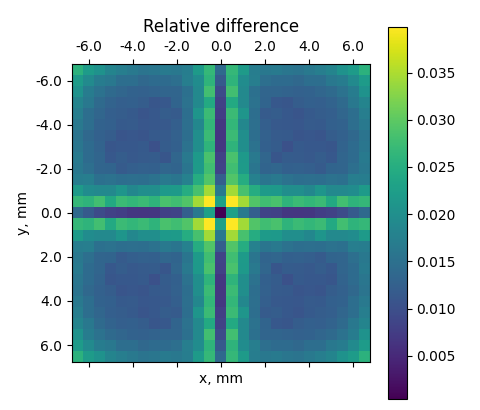}
    \caption{\label{dI_dI} Average relative difference between the KRT method (Eq. (\ref{I})) and Monte Carlo simulation in the case of fully transmitting bottom face of $14\times14\times2$ mm scintillation crystal. Reflective indexes of lateral and top faces are set to $0.95$.}
\end{figure}

Now, we consider a thinner detector of $14\times14\times2$ mm size.
In Fig. \ref{dI_dI} the relative difference averaged over photomultipliers
$\frac{1}{4}\sum_{i,j}\big|I_{ij}^{(N,N)}-I_{ij}^{(MC)}\big|/I_{ij}^{(MC)}
$ is represented for all points of scintillation crystal with a $0.5$ mm step along $x$ and $y$ axes. Here $I_{ij}^{(MC)}$ represents averaged intensity deposited in $ij$ photomultiplier in Monte Carlo simulation, $I_{ij}^{(N,N)}$ represents the intensity deposited in the $ij$ photomultiplier according to Eq. (\ref{I}) ($N=M=50$; a gamma particles penetrate the scintillation crystal exponentially, so we considered $z_0$ to be equal to its expected value: $\overline{z}_0=H-\frac{\int_0^H z \exp{(-z/l)dz}}{\int_0^{+\infty} \exp{(-z/l)dz}}$, here $l$ is the mean free path of the gamma particle). This averaged relative difference constitutes $2\%$ across the scintillator, with a peak value of about $4\%$ near the center of the scintillation crystal.
Hence, the Monte Carlo simulation results are similar to the results achieved using Eq.~(\ref{I}) in the KRT method.

\begin{figure}
\begin{minipage}{0.99\linewidth}
\raggedright{\large{a)}}

    \centering{
    \includegraphics[width=0.9\linewidth]{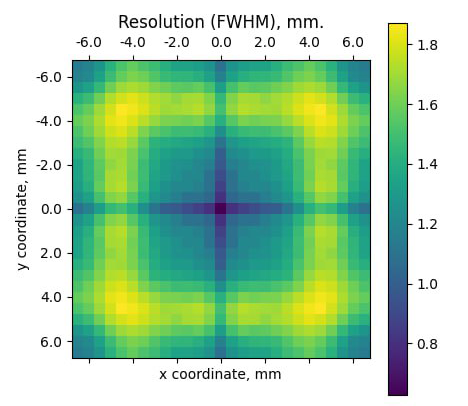}}
\end{minipage}
\begin{minipage}{0.99\linewidth}
\raggedright{\large{b)}}

    \centering{
    \includegraphics[width=0.9\linewidth]{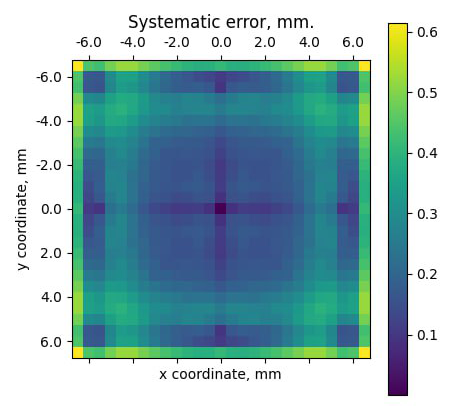}}
\end{minipage}
    \caption{Resolution (a)) and systematic error (b)) achieved using Eq.(\ref{I}) as a reference for the reconstruction algorithm. The reflective indexes of the lateral and top faces are set to $0.95$. The bottom face is fully transmitting. The scintillation crystal size is $14\times14\times2$ mm.}
\label{fwhm}
\end{figure}

Another goal was to establish the $(x,y)$ coordinate of the flash using the photomultipliers' outcome.
It was done in two ways: using the averaged results of Monte Carlo simulations and the KRT method results as reference outcomes of photomultipliers.

The idea behind averaging is simple: calculate the mean energy response over all the GEANT4 data for each coordinate $(x,y)$ being struck by gamma, and make the "$\rm{Coordinate}\ \rightarrow\ \rm{Mean\  Signals}$" mapping. Then, the procedure of coordinate reconstruction can be implemented. When a new gamma particle strikes the scintillator, it gives a vector of four numbers (energies deposited in photomultipliers). The most similar outcome, under the Euclidian norm, is found among the mapped mean signals. The $(x,y)$ coordinates of the identified most similar outcome are treated as flash coordinates.

\begin{figure}
\begin{minipage}{0.99\linewidth}
\raggedright{\large{a)}}

    \centering{
    \includegraphics[width=0.9\linewidth]{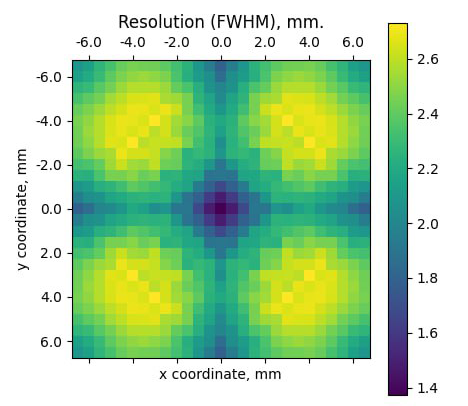}}
\end{minipage}
\begin{minipage}{0.99\linewidth}
\raggedright{\large{b)}}

    \centering{
    \includegraphics[width=0.9\linewidth]{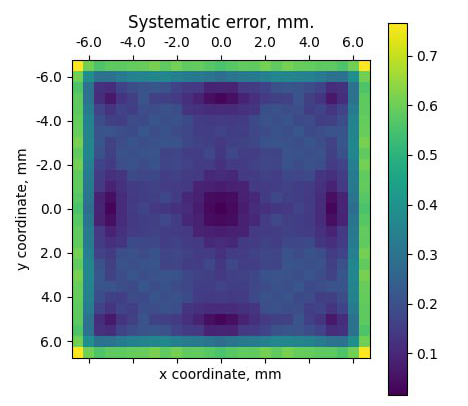}}
\end{minipage}
    \caption{Resolution (a)) and systematic error (b)) achieved using Eq.(\ref{I}) as a reference for the reconstruction algorithm. The reflective indexes of the lateral and top faces are set to $0.95$. The bottom face is fully transmitting. The scintillation crystal size is $14\times14\times7$ mm.}
\label{fwhm_7}
\end{figure}

Instead of using the averaged signal from modeling for the "$\rm{Coordinate}\ \rightarrow\ \rm{Mean\ Signals}$" mapping, one can use Eq. (\ref{I}) to calculate the photomultipliers outcome.
We implemented Eq. (\ref{I}) (with $p_{zL}=0$) with a step of 0.5 mm along the $x$ and $y$ axes at $\overline{z}_0$ height. Thus, the "$\rm{Coordinate}\ \rightarrow\ \rm{Mean\  Signals}$" mapping was achieved. After that, the procedure for the flash coordinate reconstruction described above can be implemented.

To test the resolution of both approaches, we hit each $(x,y)$ point of the scintillator with a gamma particle with energy $122$ keV $10000$ times in Geant4. If the particle produces a scintillation flash, we write down the intensity deposited in each of the photomultipliers and consider this array of intensities as truth light yield.
After that, we use this array of intensities to reconstruct the coordinates of the flash via both approaches.
Since the $x$ and $y$ coordinates are established approximately they form 2D Gaussian distribution. FWHM (full width at half maximum) was used to establish the resolution. We used $FWHM_{2d} = 2.35*\sqrt{\frac{ \sigma_{x}^{2}+\sigma_{y}^{2}}{2}}$ (as in the 1D case $FWHM = 2.35\sigma$) to determine the resolution. Systematic error was defined as follows: if $\rm{shift}_x$ is a most-likely error along the $x$-axis and $\rm{shift}_y$ along the $y$-axis, then total systematic error is $\sqrt{\frac{\rm{shift}_x^2+\rm{shift}_y^2}{2}}$.
Points $(x,y)$ were also meshed with a step of 0.5 mm.
All reflective indexes of lateral faces are $0.95$, the reflective index of the top face is $0.95$, and the reflective index of the bottom face is $0$. High values of reflective indexes are chosen to enhance the energy resolution~\cite{overwiew_anger}.

The resolution (statistical error) of the KRT method (Eq. (\ref{I})) is given in Fig.~\ref{fwhm} a) and the systematic error of this method is represented in Fig.~\ref{fwhm} b).
The maximal value of the resolution is $\sim 1.8$ mm.
The resolution is $\sim 1.2$ mm near the lateral faces of the scintillator and the center of the detector.
The maximal value of the systematic error is $\sim 0.6$ mm.
Near the center of the detector the value of the systematic error is $\sim 0.2$.
Thus, the field of view effect takes place: the resolution and positioning of the flash are better near the center of the detector.
The approach that uses Monte Carlo simulation as the reference for the flash position reconstruction shows similar results.
However, the calculation of the mean signal outcome at all points of the detector using the KRT method took approximately $0.1$ second, while the same calculation using
GEANT4 Monte Carlo simulation
took approximately an hour.

Two basic reasons cause statistic errors: the non-uniformity of the scintillation flash in an individual event (the flash might be considered uniform only on average) and the penetration depth of a gamma particle into the scintillator in an individual event.
However, the achieved resolution is at an acceptable level, being two times less than half of the photomultiplier size.
The appearance of systematic errors might be due to the non-uniqueness of the solution of Eq. (\ref{I}) or small gradients of the solution.
Indeed, if for several points in the scintillator the matrix of photomultipliers gives the same or quite a similar outcome, these points might be mixed up by the reconstruction algorithm.
However, systematic error is of the order of the step size along axes, thus it might be neglected.

For the detector with a larger crystal size of $14\times14\times7$ mm, results are presented in Fig. \ref{fwhm_7}.
Generally, they repeat the results for the crystal of $14\times14\times2$ mm size presented in Fig. \ref{fwhm}.
The main difference is that the absolute value of the resolution is higher: $2.6$ mm.
Also, the area of the detector, representing peak resolution compared to the thinner crystal, is broader.
Systematic error is $\sim 0.2$ mm overall in the crystal.
Both resolution and systematic error are at an acceptable level.

\begin{figure}
    \centering
    \includegraphics[width=0.9\linewidth]{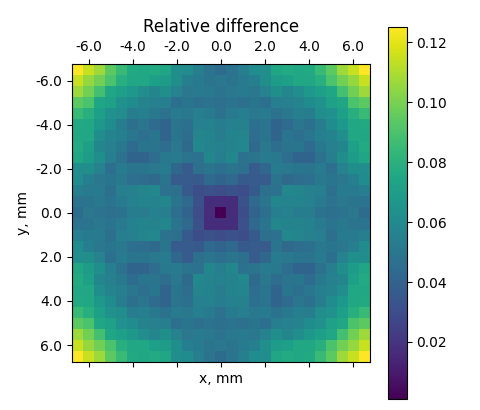}
    \caption{\label{dI_dI_IR} Average relative difference between the KRT method (Eq. (\ref{Dug})) and Monte Carlo simulation in the case of the scintillation crystal of $14\times14\times7$ mm size with $1.9$ refractive index connected via a $0.1$ mm layer of medium with a $1.5$ refractive index to the matrix of photomultipliers with a reflective index of $0.1$. Reflective indexes of the lateral and top faces are set to $0$.}
\end{figure}

The worsening of the resolution in the case of a higher crystal should be connected to the smoother gradient map of intensities deposited in the photomultipliers compared to the case of a $14\times 14\times2$ mm crystal.
Indeed, when a flash happens near the bottom surface of the scintillation crystal, a 1 mm change in the flash position can cause a sufficient change in the spherical angles occupied by each of the photomultipliers, especially when the flash happens near the border of a photomultiplier.
When flash happens at a sufficient height, a 1 mm change in the flash position may not cause a sufficient change in the mentioned spherical angles.
The flash happens approximately at the $\overline{z}_0$ height on average, which is proportional to $H$.
Thus, while the crystal is getting higher, the resolution can decrease.

\subsection{Simulation and flash coordinates reconstruction in the case of a reflecting bottom surface}\label{SFRBF}

In this section, we perform a similar analysis as in Section \ref{SFTBF}, but for the case of a reflecting bottom surface of the scintillation crystal.
We show a good agreement between the KRT method and Monte Carlo simulation with $2-3$ mm resolution for the flash coordinates reconstruction.

We consider here a scintillation crystal of $14\times14\times7$ mm size (refractive index $1.9$), with four $7\times 7$ mm photomultipliers attached to its bottom face via a thin layer of connecting medium ($0.1$ mm) with refractive index $1.5$.
Reflective indexes of the lateral and top faces are set to zero.
Reflective index of detecting face is set to $0.1$.

\begin{figure}
\begin{minipage}{0.99\linewidth}
\raggedright{\large{a)}}

    \centering{
    \includegraphics[width=0.9\linewidth]{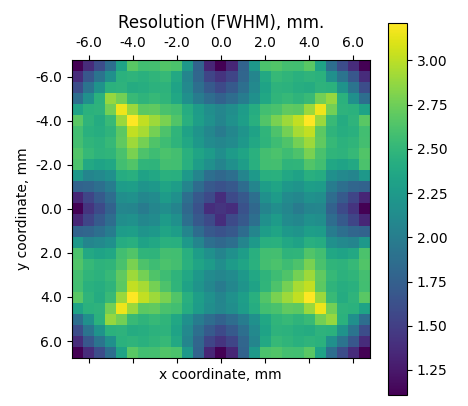}}
\end{minipage}
\begin{minipage}{0.99\linewidth}
\raggedright{\large{b)}}

    \centering{
    \includegraphics[width=0.9\linewidth]{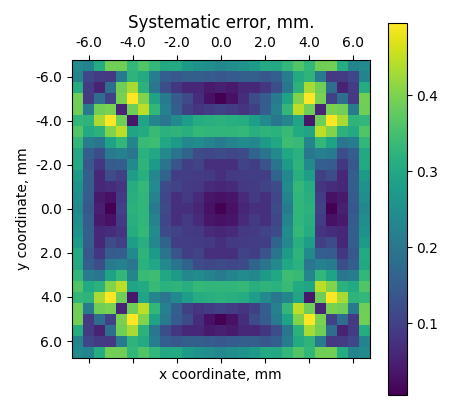}}
\end{minipage}
    \caption{Resolution (a)) and systematic error (b)) achieved using Eq.(\ref{Dug}) as a reference for the flash coordinates reconstruction algorithm. The reflective indexes of the lateral and top faces are set to $0$. The bottom face of the $14\times14\times7$ mm scintillation crystal (refractive index $1.9$) is connected within a $0.1$ mm layer of connecting medium (refractive index $1.5$) to the matrix of photomultipliers with reflective index $0.1$.}
\label{fwhm_IR_anal}
\end{figure}

In Fig. \ref{dI_dI_IR} the relative difference in intensities achieved via the KRT method and Monte Carlo simulation is depicted.
The relative difference in intensities was calculated similarly to Fig. \ref{dI_dI}.
It is seen that inside the central region, this relative difference is approximately $6\%$.
Hence, inside this region, both analytical and Monte Carlo calculations of intensities deposited in the photomultipliers show similar results.
A high relative difference near the corners of the detector is connected to the small overall intensity deposited in the opposite photomultiplier when a flash happens near the corner.

\begin{figure}
\begin{minipage}{0.99\linewidth}
\raggedright{\large{a)}}

    \centering{
    \includegraphics[width=0.9\linewidth]{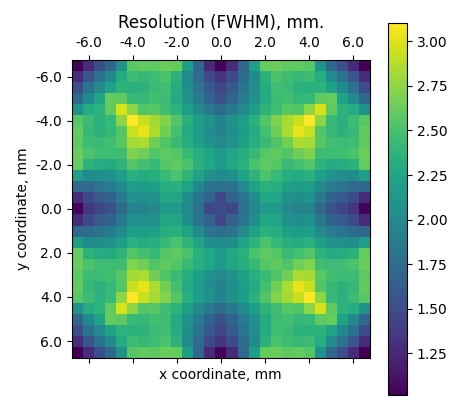}}
\end{minipage}
\begin{minipage}{0.99\linewidth}
\raggedright{\large{b)}}

    \centering{
    \includegraphics[width=0.9\linewidth]{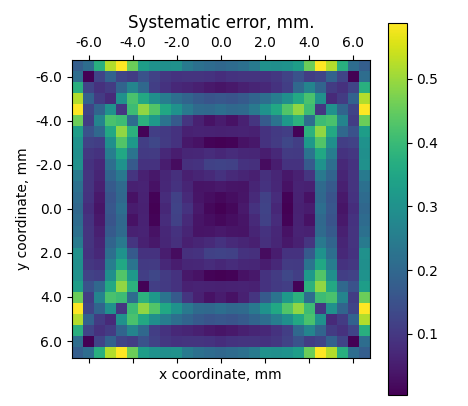}}
\end{minipage}
    \caption{Resolution (a)) and systematic error (b)) achieved using Monte Carlo simulation as a reference for the flash coordinates reconstruction algorithm. The parameters of the detector are the same as in Fig. \ref{fwhm_IR_anal}.}
\label{fwhm_IR_MC}
\end{figure}

Resolution and systematic error of flash coordinates reconstruction using Eq. (\ref{Dug}) as a reference are represented in Fig. \ref{fwhm_IR_anal}.
Resolution and systematic error of flash coordinates reconstruction using Monte Carlo simulation as a reference are represented in Fig. \ref{fwhm_IR_MC}.
Both approaches show similar resolutions of $1.5-3.0$ mm and systematic errors of $0.1-0.5$ mm.
Both approaches exhibit a field of view effect with a $0.1$ mm systematic error and a $1.5$ mm resolution near the center of the scintillation crystal.
Thus, resolution and systematic error in both approaches are at an acceptable level.

\section{Discussion and Conclusion}\label{Conc_sec}
In this article, we derived approximate analytical solutions for the calculation of intensities deposited in the matrix of photomultipliers attached to the continuous scintillator, with scintillator faces considered to be specular reflectors with reflective indexes $\sim 1$ or $0$.
The solutions were achieved by means of the kaleidoscopic ray-tracing method (KRT): detecting face was kaleidoscopically reflected over reflective faces, and intensities deposited in all reflections of detecting face were summed.
Numerical implementations of the method involve errors, which are also estimated.
The case of a fully transmitting bottom face was considered, and the general theory for the reflective bottom face was developed.

We conducted Monte Carlo simulations of scintillations inside the rectangular cuboid scintillation crystals with sizes of $14\times14\times2$ mm and $14\times14\times7$ mm using the GEANT4 \cite{GEANT4} simulation toolkit, which involves nuclear physics of the scintillation process, a non-point source of light, fluctuations in the number of photons, and asymmetry of the scintillation flash to compute the intensities deposited in the matrix of photomultipliers.
The matrix of photomultipliers consisted of four $7\times7$ mm photomultipliers.

Both scintillator sizes were examined in the case of a fully transmitting bottom face, and for the $14\times14\times7$ mm scintillator, we modeled a case when the scintillation crystal is connected via an optically less dense medium to the matrix of reflecting photomultipliers.
The last case involved total internal reflection at the crystal-connecting medium border.

We compared the results provided by the KRT method with the results provided by Monte Carlo simulations.
In all considered cases, the methods showed good agreement.

We used the intensities achieved using the KRT method and averaged Monte Carlo simulations as reference points for the scintillation flash coordinates reconstruction procedures.
Both methods experienced the "field of view" effect and showed similar spatial resolution and systematic error in all considered cases.

In the case of fully transmitting bottom face, we achieved $1-1.8$ mm resolution and $0.1-0.6$ mm systematic error for $14\times14\times2$ mm crystal and $1.5-2.6$ mm resolution and $0.1-0.6$ mm systematic error for $14\times14\times7$ mm crystal.
In the case of reflecting bottom face we achieved $1.5-3.0$ mm resolution and $0.1-0.5$ mm systematic error for $14\times14\times7$ mm crystal.
In all cases, the resolution was better than the half length of the photomultiplier.
In all cases, systematic error was approximately equal to or less than the mesh step size along the $x$ and $y$ axes ($0.5$ mm).
This means that the derived analytical solutions indeed catch all the physical aspects needed for the calculation of the intensities deposited in the photomultipliers.

The calculation of the mean signal outcome at all points of the detector in the considered cases by the KRT method took approximately $0.1$ second, while in the GEANT4 Monte Carlo simulation the same calculation took approximately an hour.
Thus, the KRT method can be implemented in real-time with the online changing of variables.
This can be useful in the optimization of scintillator detectors and in experiments for fine-tuning when some parameters of the detector and its environment change slowly (temperature, etc.).


A modification of the derived approximate solutions that takes light scattering into account was achieved.
For the considered scintillators with sizes $14\times14\times2$ mm and $14\times14\times7$ mm made from GAGG:Ge, it was shown that errors related to the light scattering are negligible.

Modifications of the KRT method, which account for the SiPM geometric factor and non-uniformity of the flash, were developed.

Limitations of the KRT method related to the scintillator shape were discussed.
The shape of the scintillator has to be one of the kaleidoscopic forms for the developed methods to be applicable.

Finally, it is important to outline areas of possible application of the developed method.
The total-body PET imaging systems can benefit from a fast calibration technique \cite{moskal2016time,moskal2021simulating,vandenberghe2020state,beltrame2011ax,zhang2017quantitative, cherry2018total}.
Also, it can be used in common X-ray CT tomography \cite{evans2006monte,kalender2006x,wang2008outlook} and SPECT \cite{SPECT_AND_PET,madsen2007recent,khalil2011molecular,bailey2013evidence}.
As well, the muon mapping of underground natural resources can utilize this calibration method since detectors for its purposes are usually designed in the same manner as in total-body PET imaging \cite{bonneville2017novel,bonneville2019borehole,procureur2018muon,d2023borehole,gluyas2019passive}.
The method can be used in proton beam range monitoring for the needs of proton therapy \cite{lang2022towards,roellinghoff2014real,litzenberg1999line,pausch2020detection}.
Another application is non-invasive investigations of cultural heritage objects \cite{lehmann2005non,mannes2015combined,janssens2017non,thickett2017using}.
From a scientific point of view, considered detectors are very important due to their use in astrophysics and  particle-physics \cite{kokubun2004improvements,aprile2010liquid,longland2006nuclear,itoh20071}.
\\

\noindent\textbf{Data Availability Statement:} No Data associated in the manuscript.


\printbibliography

\section{Appendix: Integrals}
In this Appendix, we derive Eqs. (\ref{I_A}) and (\ref{F_B}). To derive Eq. (\ref{I_A}), the integral over $y$ should be computed first in Eq. (\ref{I_rec}). To do that, the indefinite integral should be considered:
\begin{align*}
    &\int\frac{dv}{(v^2+k)^{3/2}}=\int \frac{dv}{k} \left(\frac{1}{\sqrt{v^2+k}}-\frac{v^2}{(v^2+k)^{3/2}}\right),\\ \nonumber
    &\int\frac{v^2}{(v^2+k)^{3/2}}=-\frac{v}{\sqrt{v^2+k}}+\int\frac{dv}{\sqrt{v^2+k}},\\ \nonumber
    &\int\frac{dv}{(v^2+k)^{3/2}}=\frac{v}{k\sqrt{v^2+k}}+C.
\end{align*}
Thus,
\begin{align}
    &\int\limits_{y^{(L)}}^{y^{(L)}+b}\frac{dy}{ ((x-x_0)^2+(y-y_0)^2+z_0^2))^{3/2}}=\\\nonumber
    &=\frac{y^{(L)}-y_0+b}{((x-x_0)^2+z_0^2)\sqrt{(x-x_0)^2+(y^{(L)}-y_0+b)^2+z_0^2}}-\\ \nonumber
    &-\frac{y^{(L)}-y_0}{((x-x_0)^2+z_0^2)\sqrt{(x-x_0)^2+(y^{(L)}-y_0)^2+z_0^2}}.
\end{align}

Now the integral over $x$ can be computed. The corresponding indefinite integral is
\begin{align}
    &\int \frac{du}{(u^2+c)\sqrt{u^2+p}}=
    \begin{bmatrix}
    u=\sqrt{p}\tan{q}\\
    du=\sqrt{p}\frac{dq}{\cos^2 q}
    \end{bmatrix}=\\ \nonumber
    &\int \frac{dq}{(p\tan^2 q+c)\cos{q}}=
    \int \frac{dq\cot}{(p+c \cot^2 q)\sin{q}}=\\ \nonumber
    &\begin{bmatrix}
    t=\frac{1}{\sin q}\\
    dt=-\frac{\cos q}{\sin^2 q}dq=-t \cot q dq
    \end{bmatrix}=-\int\frac{dt}{p+c(t^2-1)}=\\ \nonumber
    &-\frac{1}{p-c}\int\frac{dt}{1+\left(\sqrt{\frac{c}{p-c}}t\right)^2}=\\ \nonumber
    &-\frac{1}{\sqrt{c(p-c)}}\tan^{-1}\left(\sqrt{\frac{c}{p-c}}t\right)+C=\\ \nonumber
    &-\frac{1}{\sqrt{c(p-c)}}\tan^{-1}\left(\sqrt{\frac{c}{p-c}}\frac{1}{\sin q}\right)+C=\\ \nonumber
    &-\frac{1}{\sqrt{c(p-c)}}\tan^{-1}\left(\sqrt{\frac{c}{p-c}}\sqrt{\frac{u^2+p}{u^2}}\right)+C=\\ \nonumber
    &\frac{1}{\sqrt{c(p-c)}}\tan^{-1}\left(\sqrt{\frac{p-c}{c}}\sqrt{\frac{u^2}{u^2+p}}\right)+\tilde{C}.
\end{align}

Here $u=x-x_0$, $p=(y^{(L)}-y_0)^2+z_0^2$, and $c=z_0^2$. The last equality follows from $\tan^{-1}x+\tan^{-1}\frac{1}{x}=\frac{\pi}{2}$. Thus,

\begin{gather}
    \int\limits_{x^{(L)}-x_0}^{x^{(L)}-x_0+a}\frac{du}{(u^2+c)\sqrt{u^2+p}}=\\ \nonumber
    =\frac{\tan^{-1}\left(\frac{u\sqrt{p-c}}{\sqrt{c}\sqrt{p+u^2}}\right)}{\sqrt{c}\sqrt{p-c}}\Bigg|_{x^{(L)}-x_0}^{x^{(L)}-x_0+a}.
\end{gather}
Finally, the integral from Eq. (\ref{I_rec}) equals
\begin{align}
    &I_A=\Bigg|\frac{\mathrm{sign}(y^{(L)}-y_0+b)}{4\pi}I\times\\ \nonumber
    &\Bigg[\tan^{-1}\left(\frac{(x^{(L)}-x_0+a)|y^{(L)}-y_0+b|}{z_0\sqrt{(x^{(L)}-x_0+a)^2+(y^{(L)}-y_0+b)^2+z_0^2}}\right)-\\ \nonumber
    &-\tan^{-1}\left(\frac{(x^{(L)}-x_0)|y^{(L)}-y_0+b|}{z_0\sqrt{(x^{(L)}-x_0)^2+(y^{(L)}-y_0+b)^2+z_0^2}}\right)\Bigg]-\\  \nonumber
    &-\frac{\mathrm{sign}(y^{(L)}-y_0)}{4\pi}I\times\\ \nonumber
    &\Bigg[\tan^{-1}\left(\frac{(x^{(L)}-x_0+a)|y^{(L)}-y_0|}{z_0\sqrt{(x^{(L)}-x_0+a)^2+(y^{(L)}-y_0)^2+z_0^2}}\right)-\\ \nonumber
    &-\tan^{-1}\left(\frac{(x^{(L)}-x_0)|y^{(L)}-y_0|}{z_0\sqrt{(x^{(L)}-x_0)^2+(y^{(L)}-y_0)^2+z_0^2}}\right)\Bigg]\Bigg|.
\end{align}

The Eq. (\ref{F_B}) can be derived in the same manner. First, the integral over $y$ should be computed in Eq. (\ref{Dug})
\begin{gather}\label{Zerr}
\iint_{B}\frac{dxdy}{((x-x_0)^2+(y-y_0)^2+z_0^2)^{3/2}}=\\ \nonumber
=\int_{x_1}^{x_2}\int_{y_0+L}^{y_0+\sqrt{R_{cr}^2-(x-x_0)^2}} \frac{dxdy}{((x-x_0)^2+(y-y_0)^2+z_0^2)^{3/2}}=\\ \nonumber
=\begin{bmatrix}
    \Tilde{x}=x-x_0\\
    \Tilde{y}=y-y_0
\end{bmatrix}
=\int_{x_1-x_0}^{x_2-x_0}\int_{L}^{\sqrt{R_{cr}^2-\tilde{x}^2}} \frac{d\tilde{x} d\tilde{y}}{(\tilde{x}^2+\tilde{y}^2+z_0^2)^{3/2}}=
\end{gather}
\begin{gather}
\nonumber
    =\int_{x_1-x_0}^{x_2-x_0}d\tilde{x}\frac{\sqrt{R_{cr}^2-\tilde{x}^2}}{(\tilde{x}^2+z_0^2)\sqrt{R_{cr}^2+z_0^2}}-\\ \nonumber
    -\int_{x_1-x_0}^{x_2-x_0}d\tilde{x}\frac{L}{(\tilde{x}^2+z_0^2)\sqrt{\tilde{x}^2+L^2+z_0^2}}.
\end{gather}
The second integral is already computed above
\begin{gather}\label{Secc}
L\int\limits_{x_1-x_0}^{x_2-x_0}\frac{du}{(u^2+c)\sqrt{u^2+p}}=\\ \nonumber
=L\frac{\tan^{-1}\left(\frac{u\sqrt{p-c}}{\sqrt{c}\sqrt{p+u^2}}\right)}{\sqrt{c}\sqrt{p-c}}\Bigg|_{x_1-x_0}^{x_2-x_0}=\\ \nonumber
=\frac{\tan^{-1}\left(\frac{uL}{z_0\sqrt{L^2+z_0^2+u^2}}\right)}{z_0}\Bigg|_{x_1-x_0}^{x_2-x_0}.
\end{gather}

The first integral can be computed as follows:
\begin{gather}\label{Firr}
    \int\limits_{x_1-x_0}^{x_2-x_0}du \frac{\sqrt{R_{cr}^2-u^2}}{(u^2+z_0^2)\sqrt{R_{cr}^2+z_0^2}}=\\ \nonumber
    =\frac{1}{\sqrt{R_{cr}^2+z_0^2}}\Bigg(\frac{\sqrt{R_{cr}^2+z_0^2}} {z_0}\mathrm{tan^{-1}}\left(\frac{u\sqrt{R_{cr}^2+z_0^2}}{z_0\sqrt{R_{cr}^2-u^2}}\right)-\\ \nonumber
    -\mathrm{tan^{-1}}\left(\frac{u}{\sqrt{R_{cr}^2-u^2}}\right)\Bigg)\Bigg|_{x_1-x_0}^{x_2-x_0}.
\end{gather}
Substituting Eq. (\ref{Firr}) and Eq. (\ref{Secc}) in Eq. (\ref{Zerr}), we get Eq. (\ref{F_B}) for the function $F(u,R_{cr},z_0,L)$.

\end{document}